\documentclass{aa}

\usepackage[T1]{fontenc}
\usepackage[varg]{txfonts}
\usepackage{mathptmx}
\usepackage{amsmath}
\usepackage{amssymb}
\usepackage{amstext}
\usepackage{amsfonts}
\usepackage{mathrsfs}

\usepackage{graphicx}

\bibpunct{(}{)}{;}{a}{}{,}

\newcommand{\dust}{\textrm{d}}
\newcommand{\gas}{\textrm{g}}

\newcommand{\cm}{\textrm{cm}}
\newcommand{\au}{\textrm{au}}

\newcommand{\St}{\textrm{St}}
\newcommand{\K}{\textrm{K}}
\newcommand{\g}{\textrm{g}}

\usepackage{lscape}

\usepackage{pifont}
\newcommand{\cmark}{\ding{51}}
\newcommand{\xmark}{\ding{55}}

\usepackage{xcolor}

\begin{document}

\title{Thousands of planetesimals: Simulating the streaming instability in very large computational domains}
\author{Urs Sch\"afer\inst{\ref{Globe}}
\and Anders Johansen{\inst{\ref{Globe},\ref{Lund}}}
\and Troels Haugb{\o}lle\inst{\ref{NBI}}
\and {\AA}ke Nordlund\inst{\ref{NBI}}}
\institute{Centre for Star and Planet Formation, Globe Institute, University of Copenhagen, {\O}ster Voldgade 5-7, 1350 Copenhagen, Denmark, \email{urs.schafer@sund.ku.dk}\label{Globe}
\and Lund Observatory, Department of Astronomy and Theoretical Physics, Lund University, Box 43, 22100 Lund, Sweden\label{Lund}
\and  Niels Bohr Institute, University of Copenhagen, {\O}ster Voldgade 5-7, 1350 Copenhagen, Denmark\label{NBI}}
\date{}

\abstract{The streaming instability is a mechanism whereby pebble-sized particles in protoplanetary discs spontaneously come together in dense filaments, which collapse gravitationally to form planetesimals upon reaching the Roche density. The extent of the filaments along the orbital direction is nevertheless poorly characterised, due to a focus in the literature on small simulation domains where the behaviour of the streaming instability on large scales cannot be determined. We present here computer simulations of the streaming instability in boxes with side lengths up to~$6.4$ scale heights in the plane. This is~$32$ times larger than typically considered simulation domains and nearly a factor~$1,000$ times the volume. We show that the azimuthal extent of filaments in the non-linear state of the streaming instability is limited to approximately one gas scale height. The streaming instability will therefore not transform the pebble density field into axisymmetric rings; rather the non-linear state of the streaming instability appears as a complex structure of loosely connected filaments. Including the self-gravity of the pebbles, our simulations form up to~$4,000$ planetesimals. This allows us to probe the high-mass end of the initial mass function of planetesimals with much higher statistical confidence than previously. We find that this end is well-described by a steep exponential tapering. Since the resolution of our simulations is moderate -- a necessary trade-off given the large domains -- the mass distribution is incomplete at the low-mass end. When putting comparatively less weight on the numbers at low masses, at intermediate masses we nevertheless reproduce the power-law shape of the distribution established in previous studies.}

\titlerunning{Thousands of Planetesimals}
\authorrunning{U. Sch{\"a}fer, A. Johansen, {\AA}. Nordlund and T. Haugb{\o}lle}
\maketitle

\section{Introduction}
The streaming instability arises from the radial drift of particles in a protoplanetary disc due to the self-shielding of the drag force that couples solid particles and gas \citep{YoudinGoodman2005,JohansenYoudin2007}. In its linear phase, any initial density perturbations are amplified exponentially with time. The density of pebble-sized solids peaks at values several orders of magnitude above the gas density in the non-linear phase of the instability \citep{Johansen+etal2009, BaiStone2010b}, with high-resolution simulations reaching up to 10,000 times the gas density \citep{Johansen+etal2015}. This is one or two orders of magnitude above the Roche density, beyond which the tidal force from the star can no longer prevent gravitational contraction. Indeed, computer simulations including the gravity between the pebbles demonstrate that the non-linear evolution of the streaming instability leads to the formation of planetesimals with a characteristic size scale of 100 km \citep{Johansen+etal2007,Johansen+etal2015,Simon+etal2016,Simon+etal2017}. The characteristic size scale agrees well with the largest bodies in the asteroid belt and in the Kuiper belt \citep{Bottke+etal2005,Kavelaars+etal2021}. The specific size outcome does depend significantly on model parameters such as the distance from the star and the column density of the pebbles \citep{Liu+etal2020}. The streaming instability has nevertheless emerged during the last decade as the most promising mechanism to explain the formation of planetesimals.

The streaming instability displays positive growth rates for any combination of local particle mass-loading and particle size \citep{YoudinGoodman2005}. The particle size is represented by its Stokes number~$\St$ in the non-dimensionalisation of the dynamical problem, with~$\St$ proportional to particle size in the Epstein drag regime. Likely values of~$\St$ lie in the range between $10^{-3}$ and $10^{-1}$, depending on the efficiency of collisional sticking \citep{Birnstiel+etal2011}. The non-linear phase nevertheless shows a dramatic transition from an undulating, coherent mid-plane wave when the ratio of the pebble to gas surface density is below a threshold of approximately~$0.01$ \citep{Johansen+etal2007, BaiStone2010b} to the emergence of extremely dense filaments at higher surface density ratios and larger Stokes numbers \citep{Johansen+etal2009,Carrera+etal2015,Yang+etal2017,LiYoudin2021}. Growth of ice-covered particles by re-condensation of volatiles has therefore been proposed to lead to formation of the first planetesimal belts outside ice lines of volatile molecules such as H$_2$O \citep{Schoonenberg+etal2017,Drazkowska+etal2017,Ros+etal2019,RosJohansen2024}.

The interaction of the streaming instability with other instabilities that cause gas turbulence is more poorly understood. Adding a turbulent diffusion term to the linear stability analysis can be detrimental to the linear growth rate of the streaming instability, unless the background turbulence is very weak \citep{Umurhan+etal2020}. It is nevertheless not clear that background turbulence can simply be treated as an increased diffusion and the non-linear evolution of the streaming instability must be evaluated together with specific instabilities that cause gas turbulence. The strength of turbulence caused by the magnetorotational instability depends on the magnetic field threading the disc. The streaming instability (or more generally, the dust back-reaction drag force on the gas) amplifies particle concentrations in pressure bumps that form in turbulence caused by the magnetorotational instability when the magnetic field is weak \citep{Johansen+etal2007}, while stronger turbulence quenches the back-reaction drag force of particles on the gas \citep{Johansen+etal2011}. Even when including Ohmic dissipation that suppresses the magnetorotational instability in the mid-plane of the protoplanetary disc, density waves launched from the turbulent surface layers still have a significant dilutional effect on the development of the streaming instability in the mid-plane \citep{Yang+etal2018}. The emergence of weak pressure bumps will nevertheless also in this case act as seeds for the particle back-reaction to form pebble filaments. The same is true when considering the streaming instability in conjunction with the vertical shear instability \citep{Nelson+etal2013}. In this scenario, the streaming instability either dominates the dynamics of the dust layer or operates in an undulating layer induced by the vertical shear instability \citep{Schafer+etal2020,Schafer+etal2022}.

The size of the simulation domain has received little attention compared to the exploration of these physical conditions for the streaming instability to induce the formation of dense filaments and planetesimals. The typical size of local shearing box simulations is~$0.2$ gas scale heights \citep{Johansen+etal2009, BaiStone2010b}, a choice made from balancing the need for resolving the most unstable wavelengths of the streaming instability together with the small scales of the planetesimal-forming clumps. \citet{YangJohansen2014} consider boxes with sizes of up to~$1.6$ scale heights in order to understand the radial separation of filaments, while we examine the formation of planetesimals and their initial mass function in boxes spanning at most~$0.8$ scale heights in a previous study \citep{Schafer+etal2017}. \cite{Kowalik+etal2013} performed the only three-dimensional semi-global simulation of the streaming instability to date, but they did not take into account the vertical stellar gravity and the resulting sedimentation of pebbles towards the mid-plane.

Here, we present simulations of very large boxes, with side lengths up to~$6.4$ gas scale heights in the plane and therefore volumes up to~$1,000$ times the standard volume of streaming instability simulation domains. Our goals are twofold. Firstly, we want to understand the non-linear phase of the streaming instability and the dust dynamics it causes at all relevant scales. Secondly, we want to study the planetesimal initial mass function with unprecedentedly high planetesimal numbers.

The paper is organised as follows: In Sect.~\ref{sect:simulations}, we introduce our numerical model. This is followed by an examination of the morphology of the dust filaments caused the streaming instability in Sect.~\ref{sect:filaments}. Sect.~\ref{sect:IMF} is dedicated to the planetesimal mass distribution arising in our simulations, including its dependence on aspects of numerical modelling such as the simulation domain size, resolution, and the representation of planetesimals. We compare to the mass distribution of the cold classical Kuiper belt objects and discuss implications of our study in Sect.~\ref{sect:discussion}, and conclude in Sect.~\ref{sect:conclusion}.

\section{Simulating the streaming instability in large boxes}
\label{sect:simulations}
\begin{table*}[t]
\caption{Simulations}
\label{table:simulations}
\resizebox{\hsize}{!}{
\begin{tabular}{lccccccccccc}
\hline
\hline
Name&$L_x\times L_y\times L_z$\tablefootmark{a}&Resolution&$\St$\tablefootmark{b}&$\Pi$\tablefootmark{c}&$t_{\textrm{grav,start}}$\tablefootmark{d}&$t_{\textrm{grav, gentle}}$\tablefootmark{e}&$t_{\textrm{end}}$\tablefootmark{f}&$\rho_{\dust,\textrm{thres}}$\tablefootmark{g}&sink particle&mutual sink\\
&$[H]$&[$H^{-1}$]&&&[$P_{\textrm{K}}$]&[$P_{\textrm{K}}$]&[$P_{\textrm{K}}$]&[$\rho_{\gas,0}$]&creation criteria&particle accretion\\
\hline
\textit{run\_0.8\_320}&$0.8\times0.8\times0.2$&$320$&$\pi/10$&$0.05$&$25$&$10$&$40$&$1000$&fiducial&\xmark\\
\textit{run\_0.8\_640}&$0.8\times0.8\times0.2$&$640$&$\pi/10$&$0.05$&$25$&$10$&$40$&$1000$&fiducial&\xmark\\
\textit{run\_0.8\_640\_improved}&$0.8\times0.8\times0.2$&$640$&$\pi/10$&$0.05$&$10$&$5$&$20$&$1000$&improved&\xmark\\
\textit{run\_0.8\_1280}&$0.8\times0.8\times0.2$&$1280$&$\pi/10$&$0.05$&$10$&$5$&$20$&$1000$&fiducial&\xmark\\
\hline
\textit{run\_1.6\_320}&$1.6\times1.6\times0.2$&$320$&$\pi/10$&$0.05$&$25$&$10$&$40$&$1000$&fiducial&\xmark\\
\textit{run\_1.6\_640}&$1.6\times1.6\times0.2$&$640$&$\pi/10$&$0.05$&$25$&$10$&$40$&$1000$&fiducial&\xmark\\
\textit{run\_1.6\_640\_fastgrav}&$1.6\times1.6\times0.2$&$640$&$\pi/10$&$0.05$&$25$&$5$&$40$&$1000$&fiducial&\xmark\\
\textit{run\_1.6\_640\_slowgrav}&$1.6\times1.6\times0.2$&$640$&$\pi/10$&$0.05$&$25$&$15$&$40$&$1000$&fiducial&\xmark\\
\textit{run\_1.6\_640\_lowthres}&$1.6\times1.6\times0.2$&$640$&$\pi/10$&$0.05$&$10$&$5$&$20$&$200$&fiducial&\xmark\\
\textit{run\_1.6\_640\_improved}&$1.6\times1.6\times0.2$&$640$&$\pi/10$&$0.05$&$10$&$5$&$20$&$1000$&improved&\xmark\\
\textit{run\_1.6\_640\_mutualaccrete}&$1.6\times1.6\times0.2$&$640$&$\pi/10$&$0.05$&$25$&$10$&$40$&$1000$&fiducial&\cmark\\
\hline
\textit{run\_3.2\_320}&$3.2\times3.2\times0.2$&$320$&$\pi/10$&$0.05$&$25$&$10$&$40$&$1000$&fiducial&\xmark\\
\textit{run\_3.2\_640\_imroved}\tablefootmark{h}&$3.2\times3.2\times0.2$&$640$&$\pi/10$&$0.05$&$10$&$5$&$20$&$1000$&improved&\xmark\\
\hline
\textit{run\_6.4\_320}&$6.4\times6.4\times0.2$&$320$&$\pi/10$&$0.05$&$25$&$10$&$40$&$1000$&fiducial&\xmark\\
\textit{run\_6.4\_320\_lowStokes}&$6.4\times6.4\times0.2$&$320$&$0.1$&$0.05$&$25$&$10$&$40$&$1000$&fiducial&\xmark\\
\textit{run\_6.4\_320\_weakpresgrad}&$6.4\times6.4\times0.2$&$320$&$\pi/10$&$0.025$&$25$&$10$&$40$&$1000$&fiducial&\xmark\\
\textit{run\_6.4\_320\_strongpresgrad}&$6.4\times6.4\times0.2$&$320$&$\pi/10$&$0.1$&$25$&$10$&$40$&$1000$&fiducial&\xmark\\\end{tabular}
}
\tablefoot{
\tablefoottext{a}{Domain size in the radial, azimuthal, and vertical dimensions, where~$H$ is the gas scale height.}
\tablefoottext{b}{Stokes number of the dust particles.}
\tablefoottext{c}{Dimensionless parameter quantifying the strength of the gas pressure gradient as defined by \citet{BaiStone2010b}.}
\tablefoottext{d}{Time at which dust self-gravity is introduced, where~$P_{\textrm{K}}$ is the Keplerian orbital period.}
\tablefoottext{e}{Time over which the strength of the dust self-gravity is gradually enhanced.}
\tablefoottext{f}{Time at which simulation ends.}
\tablefoottext{g}{Threshold dust density for sink particle creation, expressed in units of the mid-plane gas density~$\rho_{\gas,0}$.}
\tablefoottext{h}{Sink particle creation begins at~\mbox{$t=14.2~P_{\K}$} instead of at~$15~P_{\K}$ because of memory limitations.}
}
\end{table*}

We employed the Pencil Code\footnote{http://pencil-code.org} \citep{PencilCodeCollaboration+etal2021} to simulate the gas, dust, and planetesimal components of protoplanetary discs, the former component on a Eulerian grid and the latter two components as Lagrangian particles. Both the stellar gravity and the drag force coupling between gas and dust were taken into account. We further utilised the Shear Advection by Fourier Interpolation scheme implemented by \citet{Johansen+etal2009a} to alleviate the time step constraint associated with the Keplerian shear.

As noted above, we simulated three-dimensional local shearing boxes \citep{GoldreichLyndenBell1965} with extents ranging from~$0.8$ to~$6.4$ gas scale heights in the radial and azimuthal dimensions and amounting to~$0.2$ scale heights in the vertical dimension. Since the streaming instability induces a dust scale height of~${\sim}1\%$ of the gas scale height \citep{YangJohansen2014, Carrera+etal2015, Schafer+etal2020}, the latter is sufficient to capture filament and planetesimal formation owing to it. The resolution of the grid was fixed at between~$320$ and~$1280$ cells per gas scale height. We applied sheared-periodic
boundary conditions \citep{Hawley+etal1995} at the radial and azimuthal
boundaries and periodic boundary conditions at the vertical
boundaries. For each of our simulations, domain size and resolution as well as all other parameters that distinguish them are listed in Table~\ref{table:simulations}.

We note that the shearing box approximation is based on the assumption that the simulation domain is sufficiently small for the disc curvature to be negligible. While the domain sizes we consider stretch this assumption, given the relatively small azimuthal extent of the dust filaments in our simulations (see Sect.~\ref{sect:filaments}), we are confident that making it does not compromise the validity of our results.

\subsection{Gas}
As we study protoplanetary discs in a local frame, we simulated gas with a constant temperature and a homogeneous initial mid-plane density. In the vertical dimension, on the other hand, the gas is initially in hydrostatic equilibrium, with the stellar gravity being balanced by a density gradient
\begin{equation}
\rho_{\gas}(z)=\rho_{\gas,0}\exp\left(-\frac{z^2}{2H^2}\right),
\end{equation}
where~$z$ is the vertical coordinate relative to the mid-plane, with the subscript~$0$ referring to this plane. In addition,~\mbox{$H=c_{\textrm{s}}/\varOmega_{\textrm{K}}$} is the gas scale height,~$c_{\textrm{s}}$ the sound speed,~$\varOmega_{\textrm{K}}=2\pi/P_{\textrm{K}}$ the Keplerian orbital frequency, and~$P_{\textrm{K}}$ the Keplerian orbital period. As detailed in \citet{YangJohansen2014}, we reformulated the equations of motion in terms of the difference between the gas density and this density stratification to ensure hydrostatic equilibrium down to machine precision.

Since a radial gas pressure gradient is required for the streaming instability to operate \citep{YoudinGoodman2005}, we imposed a pressure gradient as an additional source term in the equations of motion. We set its strength, as quantified using the dimensionless parameter established by \citet{BaiStone2010b}, to~\mbox{$\Pi=0.05$} as the default value. This is a common choice in numerical models of the streaming instability \citep[e.g.][]{YangJohansen2014, Simon+etal2016, Simon+etal2017, Schafer+etal2017}. We additionally performed one simulation each with values of~\mbox{$\Pi=0.025$} and~\mbox{$\Pi=0.1$} to test the dependence of our results on the pressure gradient strength.

\subsection{Super-particles}
To model the dust, we applied the super-particle approach that is frequently used when simulating the streaming instability \citep{YoudinJohansen2007, BaiStone2010a, Schafer+etal2020}. Every Lagrangian super-particle possesses the mass and momentum of a large number of the pebbles in protoplanetary discs -- since it is computationally infeasible to simulate every pebble individually -- but the drag force coupling to the gas of a single pebble. We used a particle block domain decomposition algorithm for load balancing \citep{Johansen+etal2011}, and the triangular-shaped cloud scheme to map between the grid and the particles \citep{HockneyEastwood1981, YoudinJohansen2007}.

The mass of the dust super-particles is determined by their total number and the initial ratio of dust to gas surface density. While the number of particles was equal to the number of grid cells, they were initialised with random position to seed the streaming instability. We chose an initially homogeneous dust-to-gas surface density ratio of~$2\%$. This value is higher than the canonical dust-to-gas ratio in the Milky Way interstellar medium, though consistent with the dust-to-gas mass ratios observed in some protoplanetary discs \citep[e.g.][]{Miotello+etal2023}. Our chosen combination of dust-to-gas surface density ratio and dust Stokes number assures the formation of dense filaments owing to the streaming instability \citep{Johansen+etal2009, Carrera+etal2015, Yang+etal2017, LiYoudin2021}.

We simulated dust with a fixed Stokes number of~\mbox{$\St=\pi/10=0.314$} or~$0.1$, with the former being the fiducial value. These Stokes numbers lie at the high end of the range of theoretically plausible values \citep{Birnstiel+etal2011}. We note that the pebbles in protoplanetary discs possess not a single size but a range of sizes, and that the behaviour of the streaming instability can depend on the number of dust sizes considered \citep{Schaffer+etal2018, Schaffer+etal2021, Krapp+etal2019, Paardekooper+etal2020, Paardekooper+etal2021, ZhuYang2021, YangZhu2021}. Nonetheless, for our choices of dust-to-gas surface density ratio and Stokes number the non-linear streaming instability behaves qualitatively similar, particularly with respect to filament formation, when modelling a dust size distribution or a fixed dust size equal to the maximum of this distribution \citep{McNally+etal2021, Schaffer+etal2021, YangZhu2021}.  

Our models include the self-gravity of the dust, which is required for planetesimals to form from the dust filaments caused by the streaming instability. We neglected the contribution of the gas to the gravitational potential because the streaming instability, while causing much stronger dust overdensities, only induces perturbations of the gas density of the order of a few percent \citep{JohansenYoudin2007, Schafer+etal2017}. To allow the dust layer to attain a state in equilibrium between sedimentation and turbulent diffusion, we introduced self-gravity into our model only after either~\mbox{$t_{\textrm{grav,start}}=10~P_{\textrm{K}}$} or~$25~P_{\textrm{K}}$. To avoid artificial sudden impulses on the particles, we further did not immediately initialise the self-gravity with its full strength, but gradually increased the strength over either~\mbox{$t_{\textrm{grav,gentle}}=5~P_{\textrm{K}}$} or~$10~P_{\textrm{K}}$. Table~\ref{table:simulations} lists the times~\mbox{$t_{\textrm{grav,start}}$} and~\mbox{$t_{\textrm{grav,gentle}}$} chosen for every simulation. 

\begin{figure}[t]
  \centering
  \includegraphics[width=\columnwidth]{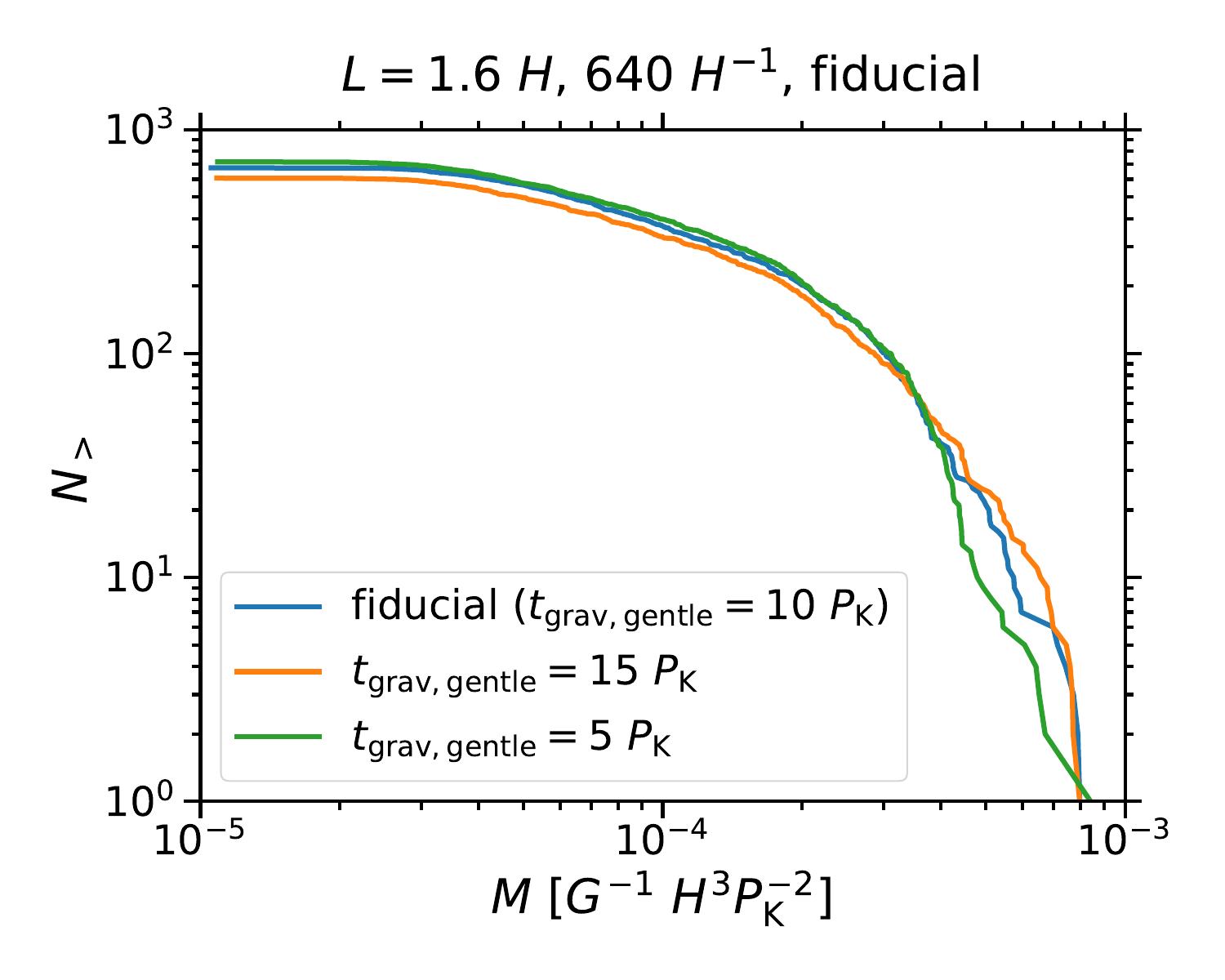}
  \caption{Cumulative number of sink particles (see Sect.~\ref{sect:sink_particles})~$N_>$ with a mass of at least~$M$ at the end of three simulations with different ramp-up times of the dust self-gravity~$t_{\textrm{grav,gentle}}$. The time at which the self-gravity is initialised is the same in all three simulations, as are all other simulation parameters. We employed the fiducial approach to sink particle formation in these simulations. The mass distributions are largely indistinguishable, apart from marginal differences at the high-mass end.}
  \label{fig:IMF_gravity_time}
\end{figure}

It is important to note that \citet{Johansen+etal2015} performed simulations similar to ours but including self-gravity from the beginning, and find the mass distribution of the emerging planetesimals to be comparable in shape to the one in our model (see Sect.~\ref{sect:IMF_fits}). Furthermore, \citet{Simon+etal2016} show that the shape of the planetesimal mass distribution is largely independent of both the strength of self-gravity and time at which it is initialised. In Figure~\ref{fig:IMF_gravity_time} we show that the mass distributions of the planetesimals forming in three simulations with different ramp-up times~$t_{\textrm{grav,gentle}}$ do not differ significantly.

This gradual enhancement in strength of the self-gravity was realised by substituting into the right-hand side of Poisson's equation
\begin{equation}
\Gamma=
\begin{cases}
0&t\leq t_{\textrm{grav,start}},\\
\frac{1}{2}\gamma\left(1-\cos\left[\frac{\pi(t-t_{\textrm{grav,start}})}{t_{\textrm{grav,gentle}}}\right]\right)&\parbox[t]{.38\columnwidth}{$t_{\textrm{grav,start}}<t<t_{\textrm{grav,start}}+t_{\textrm{grav,gentle}}$,}\\
\gamma&t\geq t_{\textrm{grav,start}}+t_{\textrm{grav,gentle}},\\
\end{cases}
\end{equation}
where the dimensionless self-gravity parameter
\begin{equation}
\gamma=\frac{4\pi G\rho_{\textrm{g,0}}}{\varOmega_{\textrm{K}}^2}=\frac{1}{\pi}.
\label{gamma}
\end{equation}
The Toomre $Q$-parameter \citep{Toomre1964} thus amounts to
\begin{equation}
Q=\frac{c_{\textrm{s}}\varOmega_{\textrm{K}}}{\pi G\Sigma_{\gas}}=\frac{4}{\sqrt{2\pi}\gamma}=5.01,    
\end{equation}
representative of a young or massive protoplanetary disc with a comparatively high disc-to-star mass ratio \citep{KratterLodato2016}; the Roche density is equal to
\begin{equation}
\rho_{\textrm{R}}=\frac{9\varOmega_{\textrm{K}}^2}{4\pi G}=28.3\rho_{\textrm{g,0}}.
\label{eq:Roche_density}
\end{equation}

We used the Keplerian orbital period~$P_{\textrm{K}}$, the gas scale height~$H$, and the mid-plane gas density~$\rho_{\gas,0}$ as the scale-free units in our model. However, these units remain independent only until self-gravity is introduced. Afterwards, the mid-plane gas density obeys
\begin{equation}
\rho_{\gas,0}=\frac{\gamma\varOmega_{\textrm{K}}^2}{4\pi G}=\frac{\pi\gamma}{GP_{\textrm{K}}^2}.
\end{equation}
The unit of mass can therefore be expressed as~\mbox{$[M]=H^3\rho_{\gas,0}=\pi\gamma~G^{-1}~H^3P_{\textrm{K}}^{-2}$}.

\subsection{Sink particles}
\label{sect:sink_particles}
We employed sink particles to represent planetesimals. In our fiducial model, one sink particle was created at the dust density maximum in every cell in which this density amounted to at least~\mbox{$\rho_{\dust,\textrm{thres}}=1000~\rho_{\gas,0}=35.3~\rho_{\textrm{R}}$}. To explore the dependence of our results on this threshold, we further considered a lower value of~\mbox{$\rho_{\dust,\textrm{thres}}=200~\rho_{\gas,0}=7.1~\rho_{\textrm{R}}$}.

Moreover -- inspired by the approaches used in models of star formation \citep[e.g.][]{Federrath+etal2010, GongOstriker2013, Haugbolle+etal2018} -- we implemented new, more sophisticated criteria for sink particle creation in the Pencil Code\footnote{These criteria can be applied using the logical flag \texttt{lsink\_create\_one\_per\_27\_cells}.}. Under these conditions, one dust super-particle was converted to a sink particle if (1) this super-particle was located in a cell where the dust density exceeded~$\rho_{\dust,\textrm{thres}}$; (2) when interpolating the gravitational potential to the super-particles, the particle represented the gravitational potential minimum inside this cell; and (3) the gravitational potential in this cell was less than those in the~$26$ neighbour cells. Considering the gravitational potential rather than the dust density for the second and third criterion is advantageous because is smoother and less affected by Poisson noise. Since the criteria involve the gravitational potential, the sink particles were generally only allowed to be created once self-gravity had attained its full strength. In one simulation their creation was permitted slightly earlier (see Table~\ref{table:simulations}), however, because the memory load due to high super-particle densities in this simulation exceeded memory limitations. 

Sink particles were allowed to accrete all super-particles, and in one of our models also all sink particles, within their accretion radius. That is, the mass and momentum of the accreted particles were added to those of the accreting sink particle, and the accreted particles were removed. The radius was chosen to be equal to one grid cell edge length. We note that this accretion thus likely is at least partly artificial because the physical accretion radius of a sink particle might be smaller than the cell size. We discuss this issue in more detail in Sect.~\ref{sect:accretion}.

\section{Large-scale pebble structure}
\label{sect:filaments}
\begin{figure}
  \centering
  \includegraphics[width=\columnwidth]{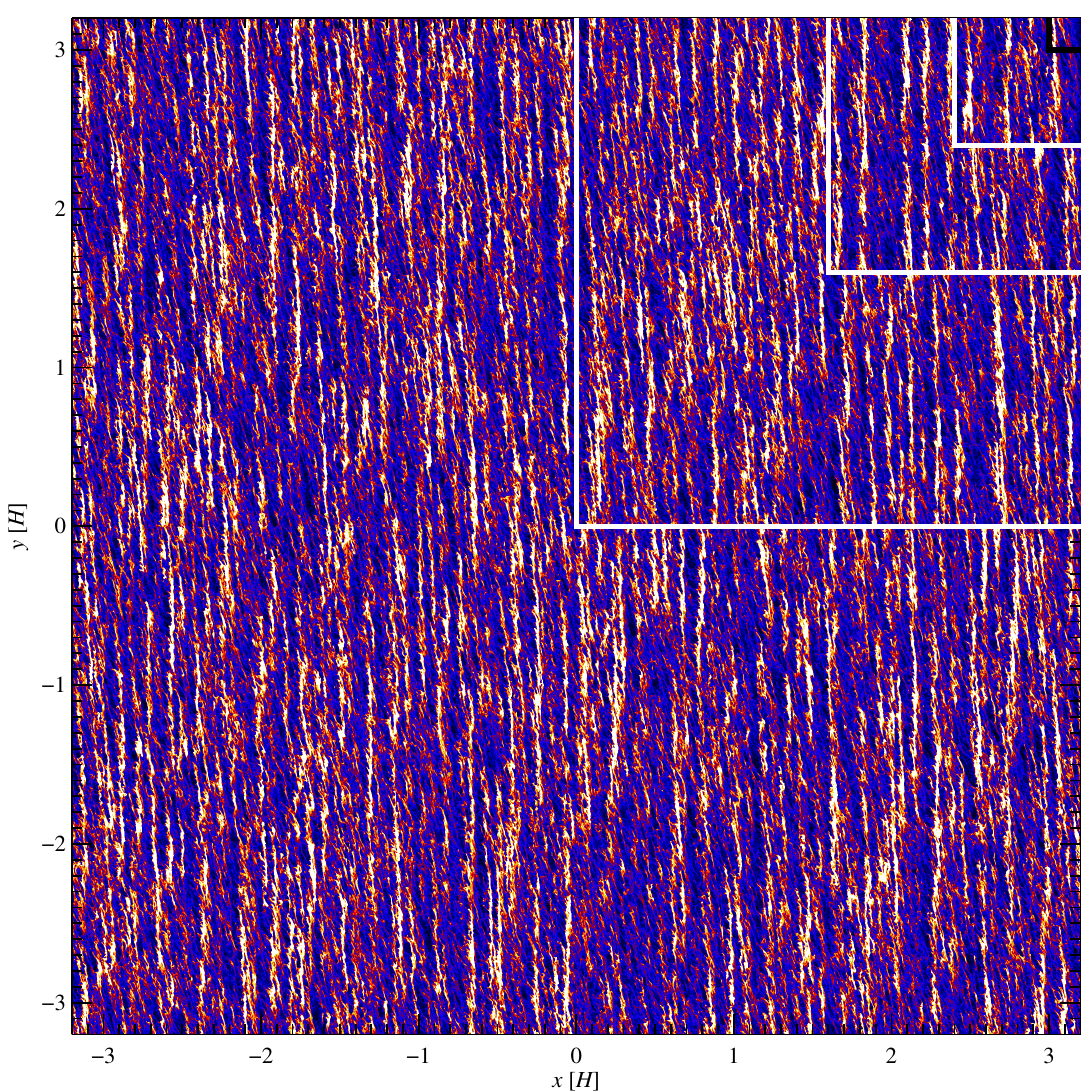}
  \caption{Ratio of the particle surface density relative to the gas surface density for simulations with box sizes~\mbox{$L=L_x=L_y=6.4,~3.2,~1.6,~0.8~H$} (white frames). The colour scale ranges from~$0.0$ to~$0.05$ (the mean surface density ratio in the simulations is~$0.02$). The nominal box size for streaming instability simulations used in most other work~(\mbox{$L=0.2~H$}) is indicated with a black frame for reference. The filamentary structure appears similar in the four box sizes.}
  \label{fig:column_ratio_comparison}
\end{figure}

We inspect the large-scale particle structure of the streaming instability turbulence in Figure~\ref{fig:column_ratio_comparison}. The figure shows overlays of the surface density ratio (particles relative to gas) for four box sizes ranging from~\mbox{$L=L_x=L_y=6.4~H$} down to~\mbox{$L=0.8~H$} at a time of~\mbox{$t=25$} orbits. The particle filaments emerging in the streaming instability turbulence are clearly not axisymmetric. Rather, the filaments appear to have azimuthal extents of up to at most one gas scale height. The particle structures in the overlaid images of smaller box sizes appear at a glance similar to the largest box size. We note that the only other authors investigating filament formation in boxes extending over more than one gas scale height in the azimuthal dimension, \citet{YangJohansen2014}, did not report the azimuthal filament extent.

\begin{figure}
    \centering
    \includegraphics{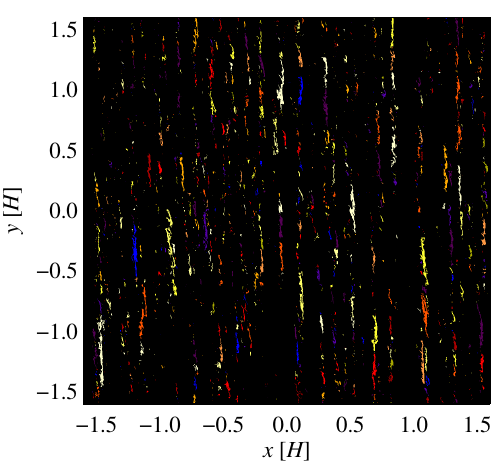}
    \caption{Mid-plane filaments captured by our filament finder algorithm for the simulation with~\mbox{$L=3.2~H$}. The colours are chosen arbitrarily to mark the different filaments. The longest filaments reach a length of order one gas scale height and appear very elongated along the azimuthal~$y$-direction.}
    \label{fig:filaments_color}
\end{figure}

We quantified the filamentary structure by analysing the dust-to-gas density ratio in the mid-plane of the simulation box\footnote{To be precise, we considered the dust-to-gas density ratio in the grid cell layer directly below the mid-plane since the mid-plane is located at the edge between two grid layers.}. We chose to consider the mid-plane density ratio rather than the surface density ratio in order to avoid contamination of the filament signal with unrelated structure above and below the mid-plane. We somewhat arbitrarily picked a dust-to-gas ratio of~$10$ as the threshold to be considered to be part of a filament. We then scanned through the grid cells that satisfy this criterion and identified connected structures. Subsequently, our algorithm merged any structures with just a single cell separation. Each dense filament was given a unique identifier number. We show the filaments identified in a simulation with~\mbox{$L=3.2~H$} in Figure~\ref{fig:filaments_color}. The largest filaments reach lengths up to a large fraction of a gas scale height and are very elongated in the azimuthal~$y$-direction.

\begin{figure}[t]
  \centering
  \includegraphics[width=\columnwidth]{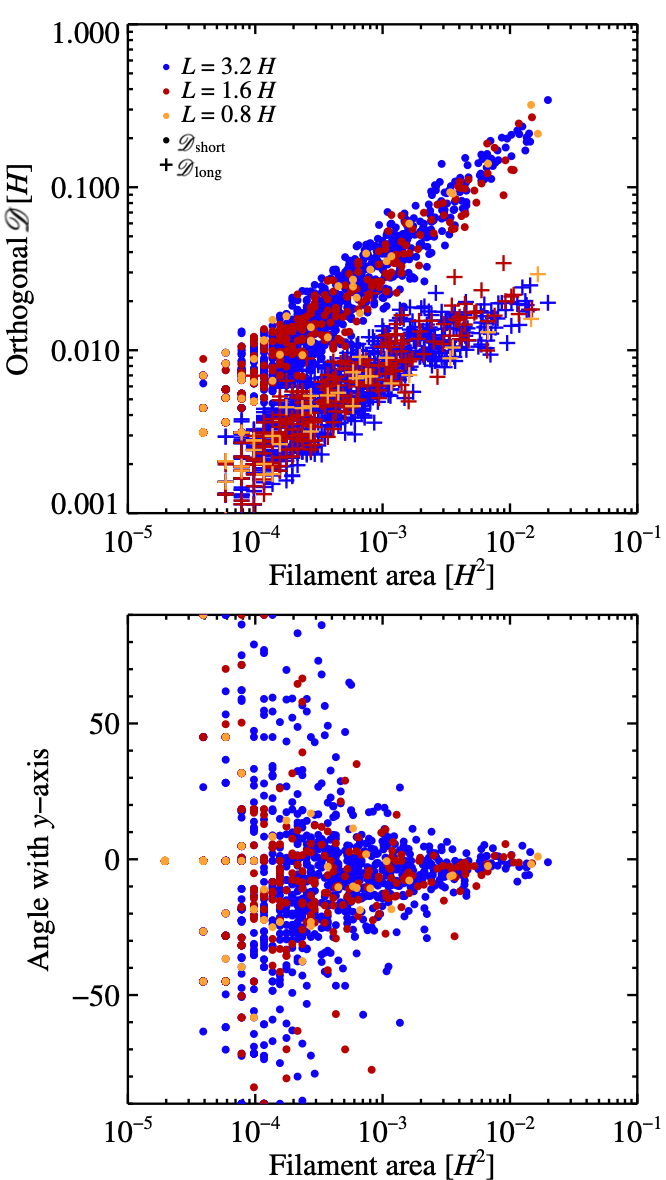}
  \caption{Ellipsoidal fits to the filaments in simulations with box sizes of~\mbox{$L=3.2~H$},~\mbox{$L=1.6~H$}, and~\mbox{$L=0.8~H$}. Top panel: Mean distance $\mathcal{D}$ relative to the short axis (circles) and the long axis (plusses). The filaments are very narrow, with a width of only~\mbox{$\mathcal{D}_{\textrm{long}}\approx0.01~H$} relative to the long axis. Their lengths relative to the short axis extend up to~$40\%$ of the gas scale height. The number of large filaments in the box with~\mbox{$L=3.2~H$} is significantly higher than in the smaller box with~\mbox{$L=1.6~H$}, due to the increased space in the large box. The smallest box with~\mbox{$L=0.8~H$} boasts two very long filaments, the same number as the box with~\mbox{$L=1.6~H$}. This could indicate an effect of the periodic boundary conditions along the~$y$-direction, which allow particle structures to connect more easily as they pass the boundary. Lower panel: Angle with the azimuthal~$y$-axis. Large filaments with areas above~$\sim$$10^{-3}~H^2$ have very small angles with the~$y$-axis.}
  \label{fig:filaments_ellipse}
\end{figure}

We quantified the geometry of the identified filaments by fitting two perpendicular axes to each filament. First, we moved the origin of the coordinate system to the centre of mass of the filament. The long axis orientation is then defined by the angle~$\theta$ with the~$y$-axis that gives the lowest mean distance measure~\mbox{$\mathcal{D}=\sum_i r_i/N$} between the filament grid cells and the line. Here,~$r_i$ is the orthogonal distance between grid cell~$i$ and the line and~$N$ is the number of cells in the filament. We define the perpendicular axis as the short axis of this ellipse fit. The short axis has the largest mean distance to the cells. These ellipse fits are compared between three simulations in Figure~\ref{fig:filaments_ellipse}. There is good convergence between the boxes with~\mbox{$L=3.2~H$} and~\mbox{$L=1.6~H$}, with far fewer large filaments formed in the smaller box. The smallest box size considered here,~\mbox{$L=0.8~H$}, nevertheless displays two large filaments with short-axis distance measures of nearly~$0.4~H$. This is likely due to the periodic boundary conditions in the $y$-direction that will artificially connect filament structures with dimensions similar to the azimuthal extent of the box.

Figure~\ref{fig:filaments_ellipse} also shows the angle of the long axis with the~$y$-axis. Small filaments have pretty random alignments, but filaments longer than~$0.02~H$ (or ~$10^{-3}~H^2$ in area) align neatly along the $y$-axis. This is expected due to the effect of the Keplerian shear that elongates any structure  evolving more slowly than the Keplerian shear time scale (essential $\varOmega_{\textrm{K}}^{-1}$).

\begin{figure*}[t]
  \centering
  \begin{minipage}{0.4\textwidth}
    \centering
    \includegraphics[width=\textwidth]{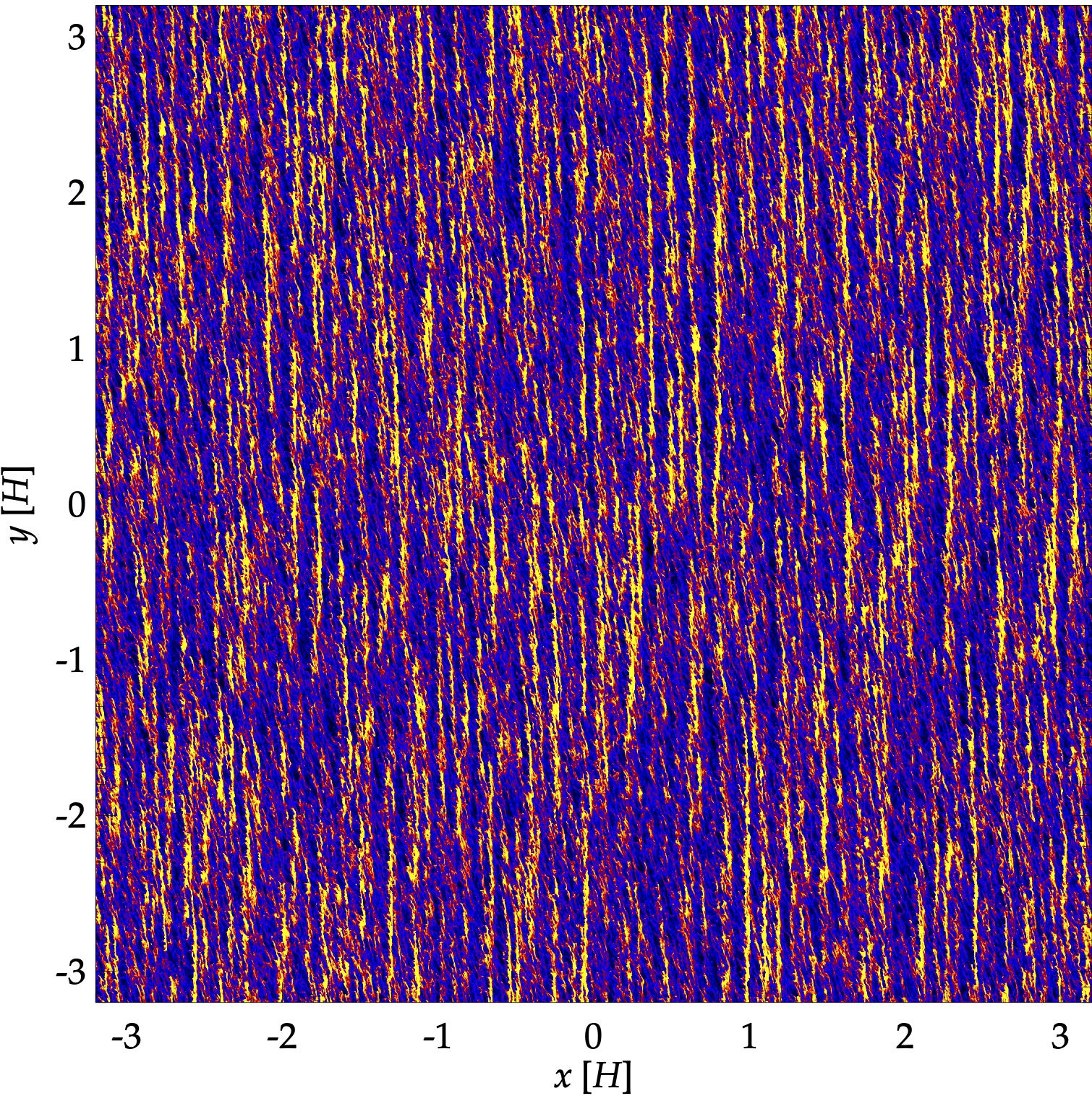} 
  \end{minipage}
  \begin{minipage}{0.419\textwidth}
    \centering
    \includegraphics[width=\textwidth]{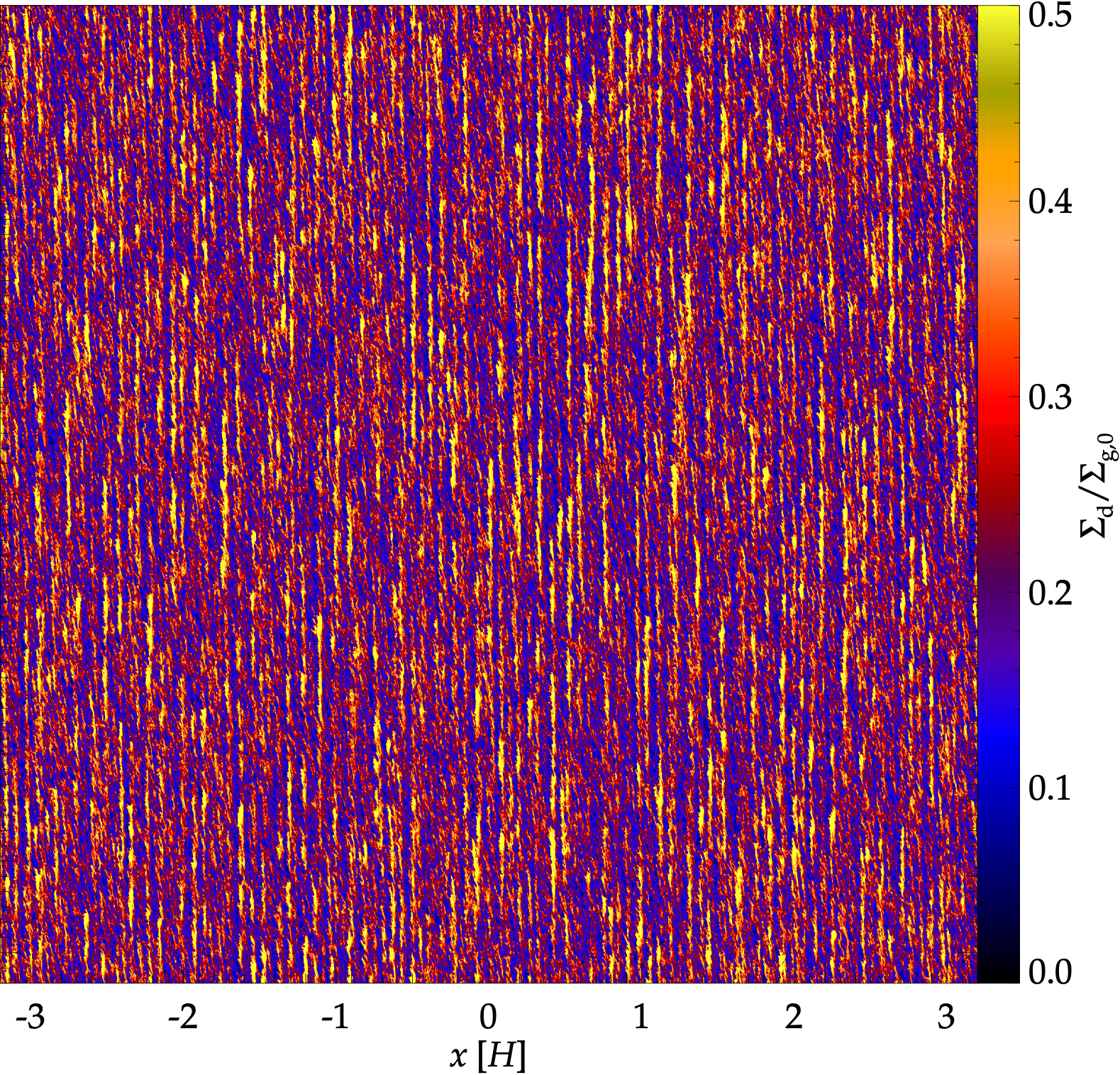} 
  \end{minipage} 
  \caption{Particle surface density in simulations with a box size of~\mbox{$L=6.4~H$} and a Stokes number of the particles of~\mbox{$\St=0.314$} (left panel) or~$0.1$ (right panel). Inspection by eye shows that the azimuthal filament size is similar in both simulations, though the particles are more dispersed in the latter one.}
  \label{fig:filaments_Stokes_number}
\end{figure*}

While particles with smaller Stokes numbers are less concentrated in filaments, this does not entail a larger azimuthal extent of these filaments. This is evident from Figure~\ref{fig:filaments_Stokes_number}, which shows that the filament size is comparable in simulations in which the Stokes number amounts to either~$0.314$ or~$0.1$.

\section{Planetesimal initial mass function}
\label{sect:IMF}
This section is dedicated to exploring the birth mass distribution, or initial mass function, of the planetesimals emerging in our simulations. To begin with, we address how simulation domain size, resolution, and our approach to using sink particles to model planetesimals affect this mass distribution. The dependences on the dust Stokes number and the strength of the radial gas pressure gradient are both examined in Appendix~\ref{sect:appendix}.

\subsection{Dependence on domain size and resolution}
\label{sect:domain_size_resolution}
\begin{figure*}[t]
  \begin{minipage}{0.49\textwidth}
    \centering
    \includegraphics[width=\textwidth]{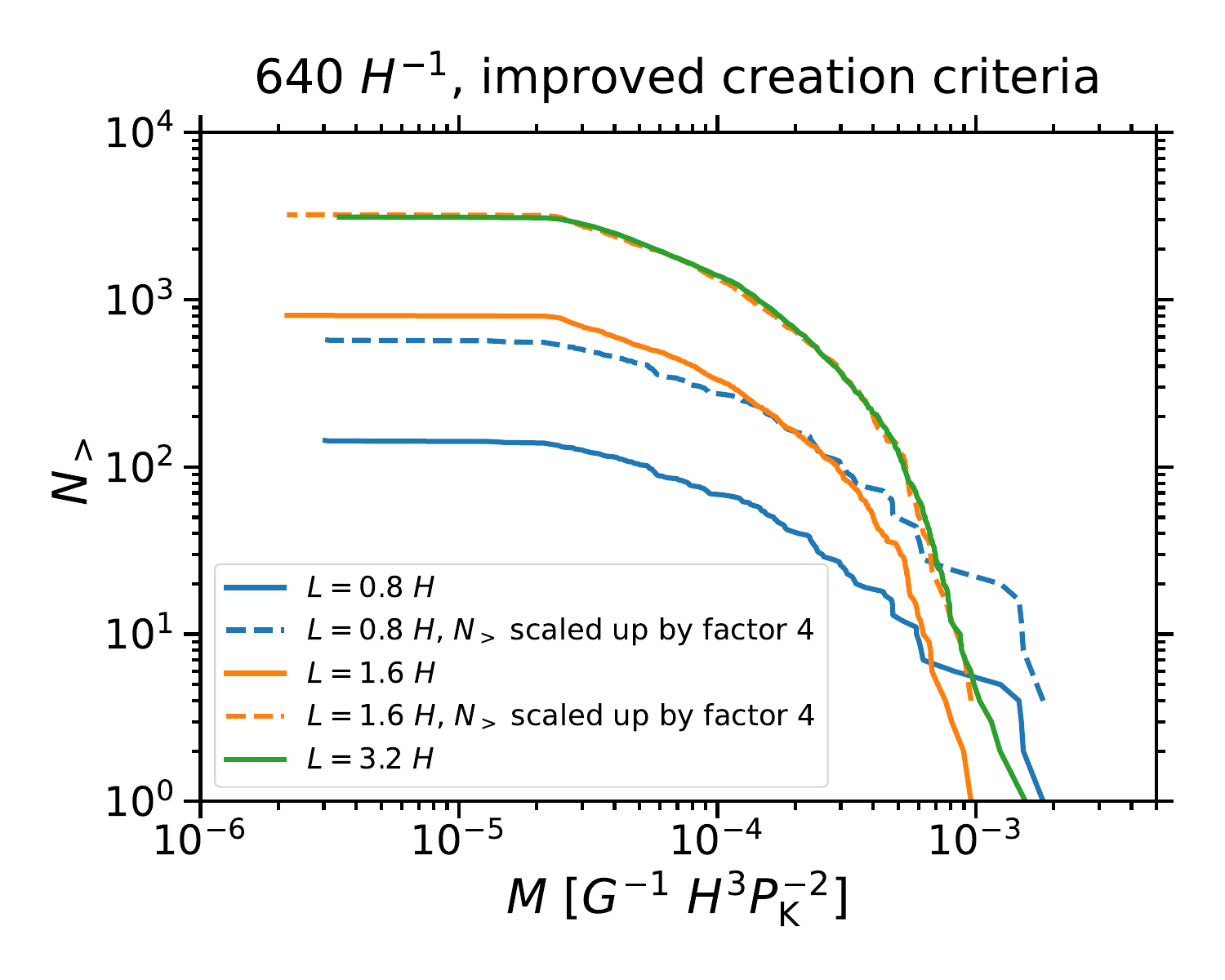} 
  \end{minipage}
  \hfill
  \begin{minipage}{0.49\textwidth}
    \centering
    \includegraphics[width=\textwidth]{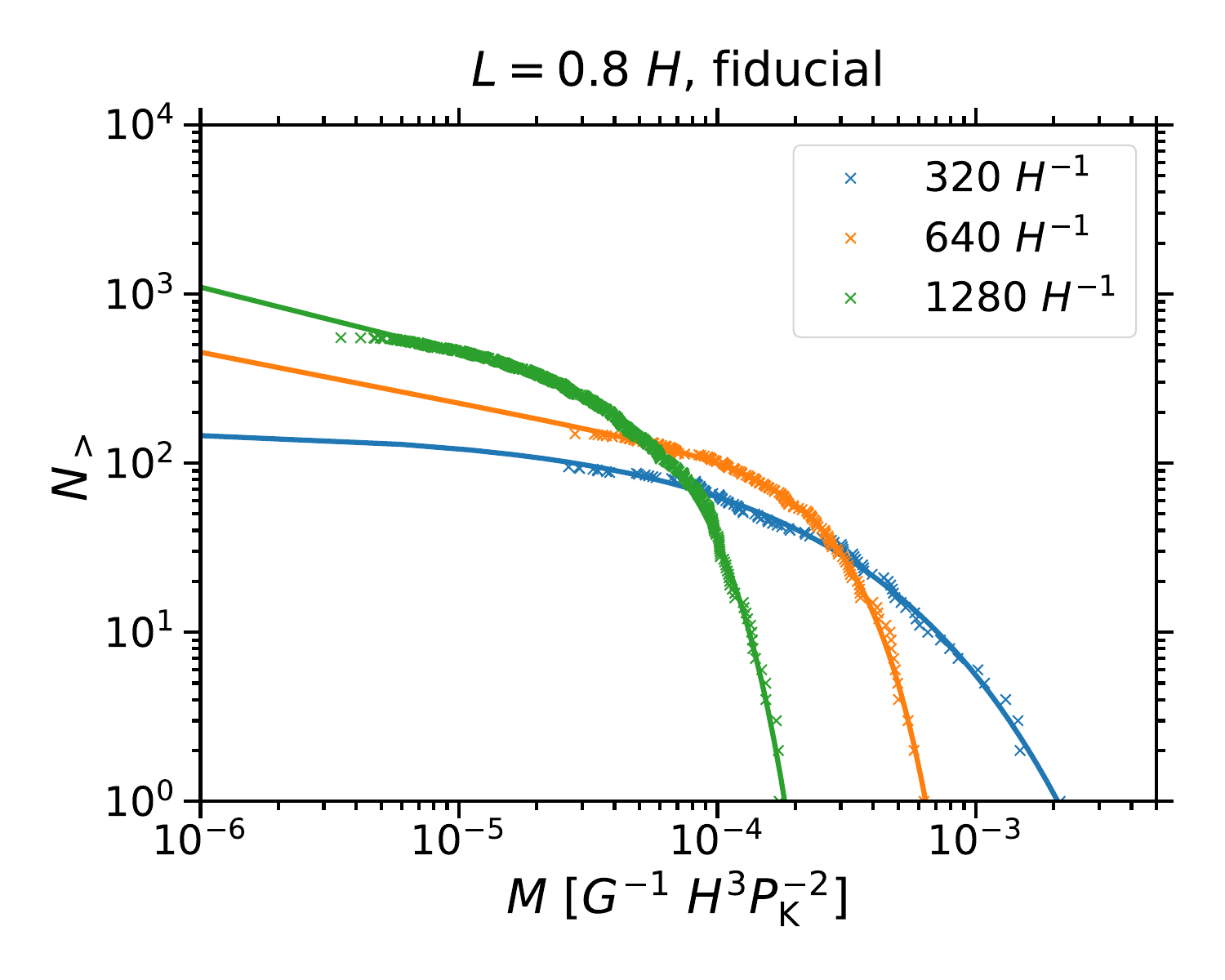} 
  \end{minipage} 
  \caption{Dependence of sink particle mass distribution on domain extent and resolution. Left panel: Domain size dependence. Solid lines represent the cumulative mass distributions at the end of simulations with radial and azimuthal domain sizes of~$L=L_x=L_y=0.8~H$ (blue line),~$L=1.6~H$ (orange line), or~$L=3.2~H$ (green line). Dashed lines, on the other hand, depict the respective cumulative numbers multiplied by a factor of~$4$, which corresponds to the difference in volume between the domains. The resolution is fixed at~$640~H^{-1}$ and the improved criteria for sink particle creation were applied in all three simulations. While the cumulative numbers in the~$L=3.2~H$-simulation are almost exactly four times higher than in the~$L=1.6~H$-simulation, for low/high masses they are more/less than four times greater in the~$L=1.6~H$-simulation than in the~$L=0.8~H$-simulation. Right panel: Resolution dependence. Cumulative mass distributions at the end of simulations with a domain extent of~$L=0.8~H$, the fiducial sink particle creation criteria, and resolutions of~$320~H^{-1}$ (blue crosses),~$640~H^{-1}$ (orange crosses), or~$1280~H^{-1}$ (green crosses). The distributions are each fitted with an exponentially tapered power law (solid lines). Both the power-law part and the exponential tapering are comparably steep for the two higher resolutions, but significantly shallower for the lowest resolution. In the latter case, the distribution is in fact represented best by only an exponential function.}
  \label{fig:IMF_domain_size_resolution}
\end{figure*}

Our simulations indicate that a domain size of~$1.6~H$ or more in the radial and azimuthal dimensions as well as a resolution of at least~$640$ grid cells per gas scale height are required for the planetesimal initial mass function to converge in shape. In the left panel of Figure~\ref{fig:IMF_domain_size_resolution}, we show the cumulative mass distribution of the planetesimals emerging in simulations with different domain extents in the plane. For any given mass, the number of planetesimals is quite precisely four times greater in the simulation with~\mbox{$L=L_x=L_y=3.2~H$} than in the one with~\mbox{$L=1.6~H$}, that is when the domain volume is quadrupled. In contrast, in the domain with~\mbox{$L=0.8~H$} more high-mass and fewer low-mass planetesimals emerge relative to the larger two domains. That is to say, the numbers at the high-mass end are less than four times larger in the~\mbox{$L=1.6~H$}-simulation than in the~\mbox{$L=0.8~H$}-simulation, but more than four times higher at the low-mass end. This is likely due to the fact that, as discussed above, the azimuthal extent of the dust filaments caused by the streaming instability typically amounts to~$1~H$. Filaments in our smaller domains thus behave as if they were axisymmetric because of the periodic boundary conditions, and planetesimals emerging from these filaments can grow more massive because their feeding zone is unbounded in the azimuthal dimension.

Figure~\ref{fig:IMF_domain_size_resolution} additionally illustrates the resolution dependence of the shape of the planetesimal mass distribution. In the right panel, the distributions in simulations with a domain size of~\mbox{$L=0.8~H$} and three resolutions are shown, each fitted with an exponentially tapered power law -- we note that performing a similar resolution study using larger domains is too demanding in terms of computational resources. The slopes of power-law part and exponential tapering are similar for resolutions of~$640$ and~$1280$ grid cells per gas scale height. In contrast, for~$320$ cells per scale height the tapering is considerably shallower while the power law vanishes, that is to say the mass distribution is best fitted with solely an exponential function.

This is because higher resolutions enable the formation of less massive planetesimals, and conversely it is not possible to constrain the power-law part of the mass distribution if the resolution is too low because of a cutoff and incompleteness at the low-mass end \citep{Johansen+etal2015, Simon+etal2016, Li+etal2019}. Less massive planetesimals form in our simulation with~$1280~H^{-1}$ than in the two lower-resolution simulations -- though, curiously, the minimum planetesimal mass in these two simulations is comparable. Similarly, the resolution study involving the highest resolutions to date conducted by \citet{Simon+etal2016} shows that the minimum mass does not converge for resolutions of up to~$2560~H^{-1}$. Nonetheless, it is encouraging that the power-law part of the planetesimal mass distribution emerges already at a resolution of~$640~H^{-1}$, and that its steepness indeed is similar to what is measured in previous studies employing higher resolutions, as we discuss in Sect.~\ref{sect:IMF_results} \citep{Johansen+etal2015, Simon+etal2016, Simon+etal2017, RucskaWadsley2021}. In contrast, the overall shallower mass distribution in our simulation with~$320~H^{-1}$ likely results from clusters of unresolved less massive planetesimals appearing as one more massive planetesimal. When examining the planetesimal birth mass distribution in what follows, we therefore disregard our simulations with domain sizes smaller than~\mbox{$L=1.6~H$} or resolutions of less than~$640~H^{-1}$.

\subsection{Sink particles}
\subsubsection{Creation}
\begin{figure}[t]
  \centering
  \includegraphics[width=\columnwidth]{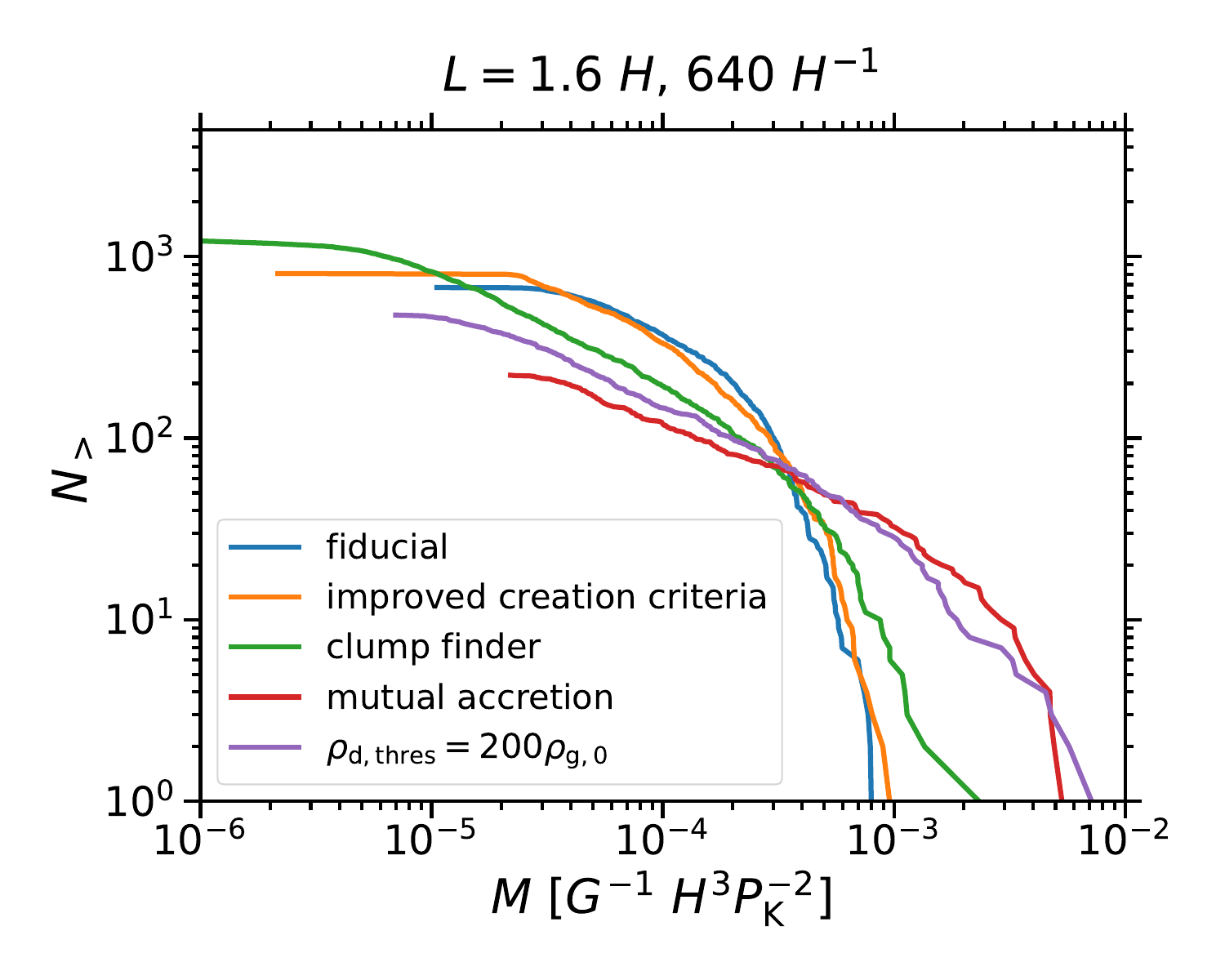}
  \caption{Cumulative mass distributions of the sink particles at the end of a simulation of our fiducial model (blue line), one with the improved sink particle creation criteria (orange line), one in which mutual sink particle accretion is permitted (red line), and one where the dust density threshold for sink particle creation is five times smaller than the fiducial value (purple line). In addition, the mass distribution of gravitationally unstable clumps identified by a clump finding algorithm is shown (green line). The distribution of sink particle masses is largely independent of whether the fiducial or the improved sink particle creation criteria are employed. However, the clump finding algorithm discovers more high-mass and fewer medium-mass clumps, while even more high-mass and fewer low-mass sink particles emerge in the simulations including sink particles mergers and a reduced dust density threshold, respectively.}
  \label{fig:IMF_sink_params}
\end{figure}

In Figure~\ref{fig:IMF_sink_params}, we show the planetesimal initial mass functions arising in our fiducial model, which includes the standard criteria for sink particle creation and excludes sink particle mergers, as well as in models with either the improved creation criteria, or with a reduced dust density threshold for creation, or with mergers. Additionally, the mass distribution of planetesimals identified by a clump finding algorithm in a model without sink particles is depicted. It is interesting to note that all these distributions intersect at a planetesimal mass of~$3\times10^{-4}~G^{-1}\,H^3P_{\rm K}^{-2}$, though we cannot provide an explanation for this finding and therefore consider this to be a coincidence.

Encouragingly, applying the fiducial or the improved sink particle creation criteria results in largely comparable initial mass functions, with only marginally more high-mass and fewer medium-mass planetesimals forming in the simulation with the latter criteria. Additionally, low-mass planetesimals are slightly more numerous and the minimum planetesimal mass smaller, an unexpected finding when considering that the improved criteria are stricter. This most probably is a consequence of sink particle creation only being permitted after the dust self-gravity has attained its full strength when the improved criteria are applied, but also before when the fiducial criteria are employed. Thus, in the latter case sink particles emerging during the time of self-gravity ramp-up can accrete dust clumps that in the former case would form sink particles themselves when the ramp-up is completed.

\subsubsection{Mutual accretion}
\label{sect:accretion}
From Figure~\ref{fig:IMF_sink_params}, it can further be seen that substantially more high-mass and fewer low-mass planetesimals emerge if mutual sink particle accretion is permitted. We note, though, that this accretion is likely to a large extent artificial (see also the discussion in Sect.~3.2 of \citealt{Schafer+etal2017}).

In principle, one planetesimal or sink particle should emerge from every gravitationally unstable dust clump -- we cannot resolve binaries or multiples -- and mergers of sink particles forming from the same clump are thus desirable. One sink particle was created per cell when the fiducial creation criteria were applied, or one per~$27$ cells when the improved criteria were used, but unstable clumps generally encompass a greater number of cells. Nonetheless, of the~$460$ sink particles that are accreted in our simulation including mergers, only~$14$ or~$3\%$ are accreted within~$0.1$ orbital periods -- the frequency with which sink particle data in our simulations were written out -- after they emerge. That is, only a small percentage is probably associated with the merging of sink particles forming from the same clump.

\begin{figure}[t]
  \centering
  \includegraphics[width=\columnwidth]{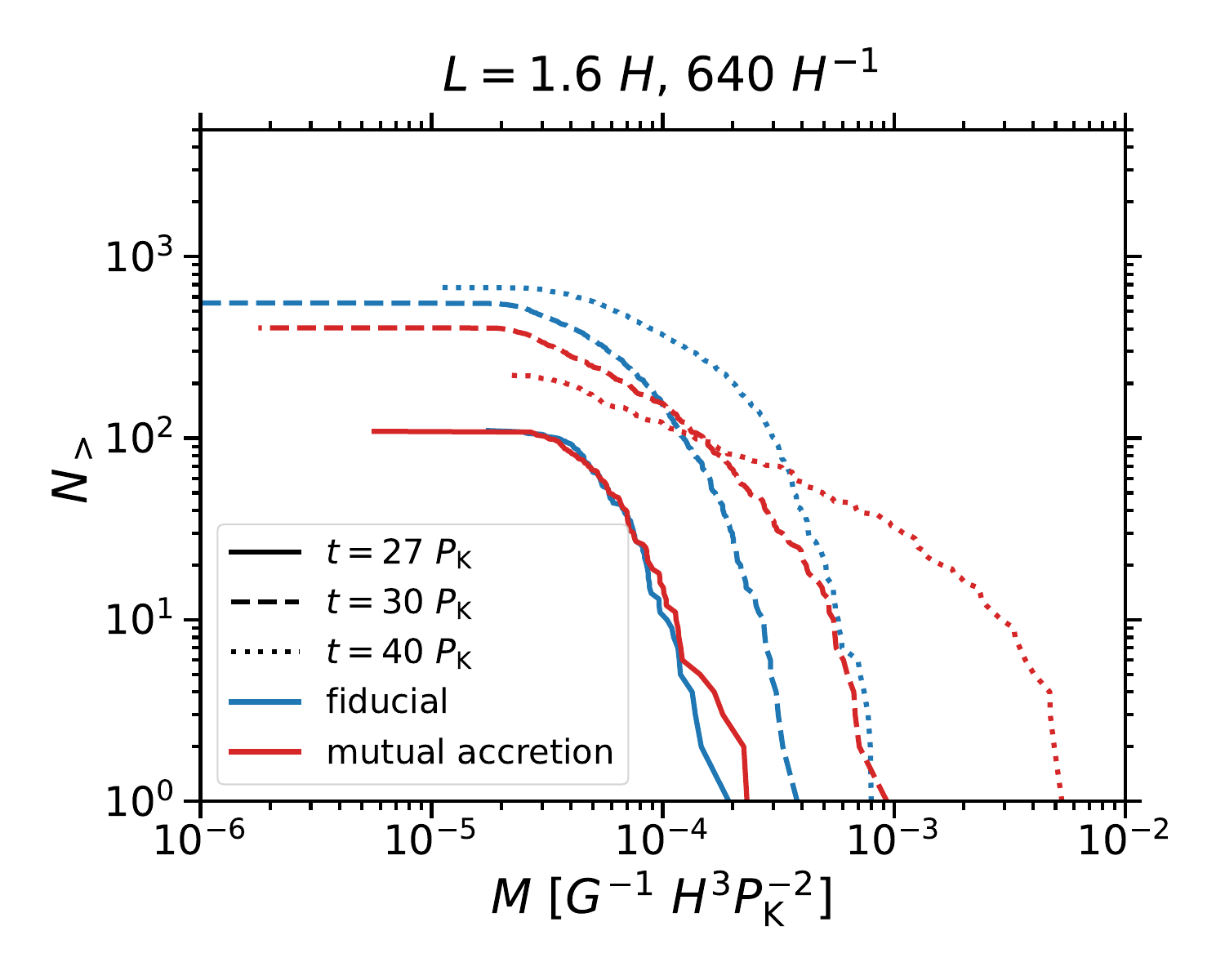}
  \caption{Cumulative mass distributions at different times in simulations excluding (blue lines) and including mutual sink particle accretion (red lines). The fiducial criteria for sink particle creation were used in both simulations. At a time of~\mbox{$t=27~P_{\textrm{K}}$} (solid lines), the distributions in the two simulations are similar. After~$30~P_{\textrm{K}}$, however, the distribution in the simulation without sink particle mergers is overall steeper than the one in the simulation with mergers (dashed lines), and this discrepancy is even stronger after~$40~P_{\textrm{K}}$ (dotted lines). The shape of the distribution in the simulation excluding mergers remains roughly the same.}
  \label{fig:IMF_time_series}
\end{figure} 

Figure~\ref{fig:IMF_time_series} depicts the evolution of the planetesimal mass distribution in two otherwise identical simulations, one each including and excluding mutual accretion, and corroborates that most mergers are likely of sink particles originating from different dust clumps. Early on, two orbits after self-gravity has been initialised, the mass distributions of the~\mbox{${\sim}100$} planetesimals that have already formed in the two simulations are comparable. Over time, however, the distribution in the simulation with mutual accretion on the whole develops to be more and more shallow compared to the distribution in the simulation without, with the shape of the latter distribution evolving comparatively little. This substantiates that sink particles are less likely to be accreted immediately by others forming from same dust clump, but more likely later on by ones that emerge from a different clump.

More generally, sink particle mergers are not resolved in our model and the majority are probably artificial. Physically, a collision occurs if the accreted sink particle is located within the maximum impact parameter of the accreting sink particle. This maximum impact parameter, taking into account gravitational focusing, can be expressed as
\begin{equation}
b_{\rm{max}}=\sqrt{(R_1+R_2)^2+\frac{2G(M_1+M_2)(R_1+R_2)}{\Delta v^2}},
\end{equation}
where~$R_i$ and~$M_i$ are radii and masses of the two sink particles and~$\Delta v$ is their relative velocity. Here, we compute radii from masses assuming spherical bodies with a solid density of~$3~\g\,\cm^{-3}$. Including only sink particles that are accreted later than~$0.1~P_{\textrm{K}}$ after their formation, and weighting by their lifetime, the average maximum impact parameter amounts to only~$4\%$ of the grid cell edge length in our simulation in which mutual accretion is allowed. The simulated accretion radius of the sink particles, on the other hand, is equal to one grid cell edge length. That is, the maximum impact parameter is on average overestimated by a factor of~$25$, and the collisional cross section by a factor of~$625$.

Nevertheless, we continue to consider our model including mutual sink particle accretion in the following analysis as an extreme case with most probably too many sink particle mergers, with the simulations excluding it as the other extreme with no mergers. 

As we show in Figure~\ref{fig:IMF_sink_params}, the planetesimal initial mass function in the simulation with a reduced dust density threshold for sink particle creation is similar to the one in the simulation with sink particle mergers. This is because this lower threshold is reached and sink particles form even before the introduction of the dust self-gravity. These early-formed sink particles then likely accrete dust clumps that would otherwise later form sink particles themselves. As self-gravity is not yet initialised, this early sink particle formation is not associated with the gravitational collapse of dust clumps. In what follows, we therefore neglect our model with the reduced density threshold.

\subsubsection{Clump finder}
While numerical studies of the streaming instability employing the Pencil Code involve sink particles to identify planetesimals \citep{Johansen+etal2015, Schafer+etal2017}, studies applying Athena \citep[e.g.][]{Simon+etal2016, Simon+etal2017, Li+etal2019, RucskaWadsley2021} rely on clump finding algorithms for identification. To bridge this gap and facilitate comparison with the latter studies as well as between our different approaches to sink particle creation and accretion, we conducted a simulation without sink particles and applied a basic clump finder at the time when self-gravity has attained its full strength. This clump finder associated planetesimals with connected structures of neighbouring cells in which the dust density exceeds the Roche density. We emphasise that the algorithms utilised by other authors to identify clumps, like PLAN \citep{Li2019}, are more sophisticated. They can, for instance, distinguish two gravitationally bound clumps even if they are bridged by cells with dust densities greater than the Roche density, and disregard clumps that are too small or not sufficiently dense to undergo gravitational collapse.

\begin{figure}[t]
  \centering
  \includegraphics[width=\columnwidth]{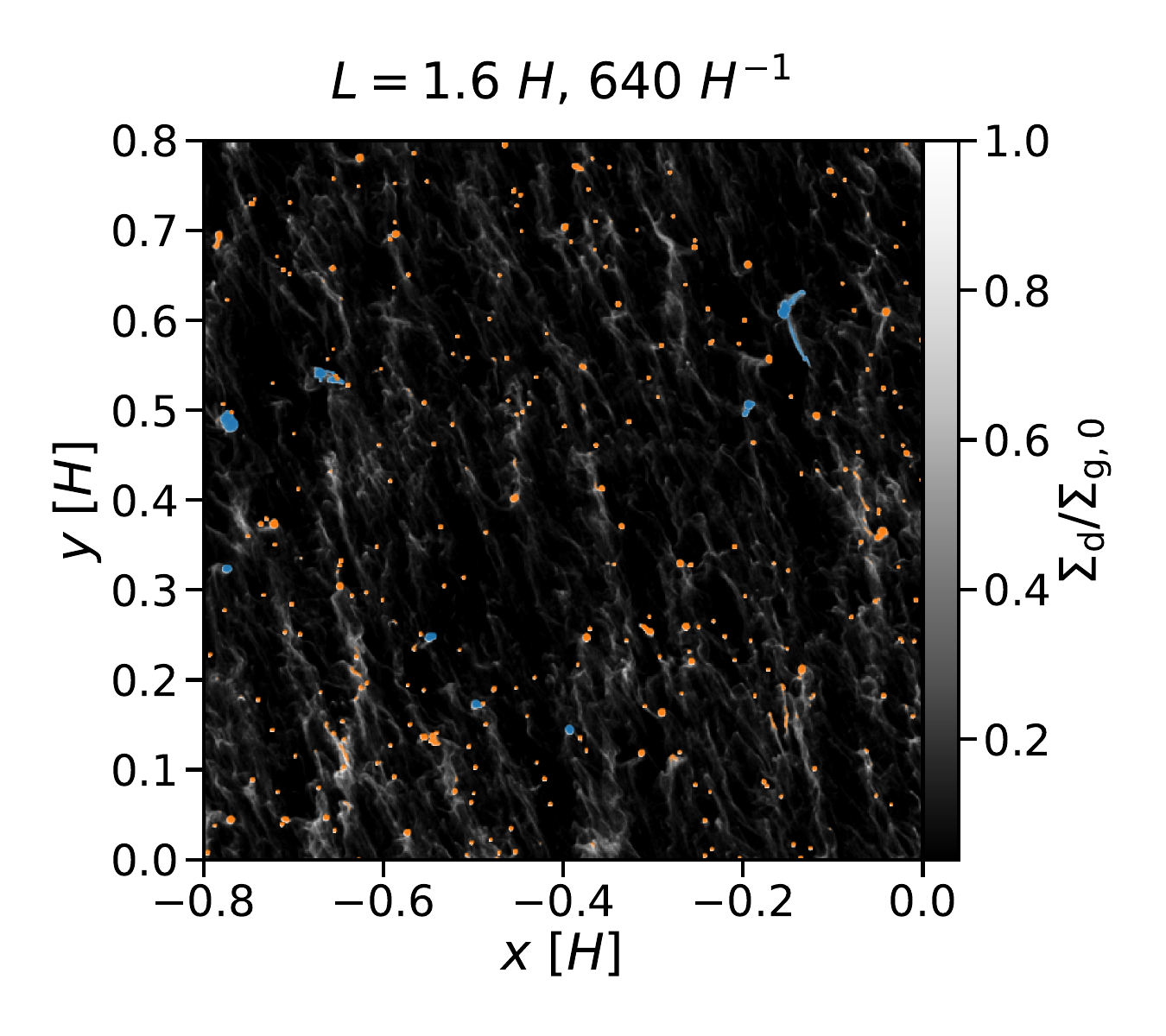}
  \caption{Dust surface density at the time when self-gravity has reached its full strength in a simulation without sink particles. The gravitationally unstable dust clumps found by the clump finding algorithm described in the main text are overplotted in blue if they are among the~$25$ most massive, and in orange otherwise. These most massive clumps, most notably one located at~\mbox{$x=-0.2~H$} and~\mbox{$y=0.6~H$}, tend to be more extended than compact structures.}
  \label{fig:clump_finder}
\end{figure}

As illustrated in Figure~\ref{fig:IMF_sink_params}, the initial mass function obtained using our rudimentary clump finder is more shallow, that is to say there is a greater number with high masses and a smaller number with medium masses, compared with the mass distributions of the sink particles in our simulations excluding mutual accretion -- though it is still steeper than the sink particle mass distribution in the simulation including mutual accretion. The reason for this is that, as one sink particle is created per cell or per~$27$ cells, often multiple sink particles would emerge from one of the gravitationally unstable clumps identified by the clump finder which encompass as many as several hundred cells.

This raises the question of whether the mass distribution of these clumps, despite being measured using a rather simplistic approach, more accurately represents the planetesimal initial mass function than the sink particle mass distribution. Figure~\ref{fig:clump_finder} shows the morphology of the clumps and indicates that the truth in all likelihood lies somewhere in the middle. The most massive clumps, which are depicted in blue in the figure, at least partly are elongated structures that in nature would possibly fragment into multiple planetesimals. 

On the other hand, a large number of clumps with very low masses is found. This is since the clump finding algorithm involves all cells with dust densities of at least the Roche density. Even clumps consisting of only one or a few cells in which the Roche density is barely exceeded, that is to say clumps that would not undergo gravitational collapse and form planetesimals, thus factor into the clump mass distribution. When fitting this mass distribution in the following, clumps with masses less than~$5\times10^{-6}~G^{-1}\,H^3P_{\rm K}^{-2}$ are thus not taken into consideration.

\subsection{Fits to the initial mass function}
\label{sect:IMF_fits}
In this section, we investigate the shape of the planetesimal mass distribution by fitting to the distributions arising in our simulations. We assume that these distributions can be described by an exponentially tapered power law, in agreement with both previous numerical studies of the streaming instability \citep{Johansen+etal2015,Schafer+etal2017,Abod+etal2019} and with the shape of the birth mass distribution of the cold classical Kuiper belt objects \citep{Kavelaars+etal2021, Napier+etal2024}.

\subsubsection{Method}
An exponentially tapered power law can be expressed as
\begin{equation}
\frac{N_>(M)}{N_{\textrm{tot}}}=\left(\frac{M}{M_{\textrm{pow}}}\right)^{-\alpha}~\exp\left[-\left(\frac{M}{M_{\textrm{exp}}}\right)^{\beta}\right],
\label{eq:tapered_power_law}
\end{equation}
where~$N_>(M)$ is the number of planetesimals with a mass equal to or exceeding~$M$ and~$N_{\textrm{tot}}$ is the total number of planetesimals. The characteristic mass scales of power law~$M_{\textrm{pow}}$ and exponential tapering~$M_{\textrm{exp}}$ as well as the exponents~$\alpha$ and~$\beta$ were all considered to be fitting parameters. This cumulative mass distribution can be converted to a differential mass distribution
\begin{equation}
\begin{split}
\frac{dN}{dM}=&-\frac{1}{M}~\left[\alpha+\beta\left(\frac{M}{M_{\textrm{exp}}}\right)^{\beta}\right]\times\\
&\left(\frac{M}{M_{\textrm{pow}}}\right)^{-\alpha}~\exp\left[-\left(\frac{M}{M_{\textrm{exp}}}\right)^{\beta}\right]
\end{split}
\end{equation}
or a cumulative size distribution
\begin{equation}
\frac{N_>(R)}{N_{\textrm{tot}}}=\left(\frac{R}{R_{\textrm{pow}}}\right)^{-3\alpha}~\exp\left[-\left(\frac{R}{R_{\textrm{exp}}}\right)^{3\beta}\right].
\end{equation}

An unprecedentedly large sample of massive planetesimals emerges in our very large domains and permits us to strongly constrain the exponential tapering. As a trade-off, the resolution of our models is limited, both in terms of the grid cell size and in terms of a cutoff and incompleteness at low planetesimal masses \citep{Johansen+etal2015, Simon+etal2016, Li+etal2019}. We can thus only put weak constraints on the power-law part of the initial mass function compared with previous studies employing smaller domains with higher resolutions.

We elected to fit the planetesimal mass distributions at the end of our simulations. On the one hand, in our previous study \citep{Schafer+etal2017} we show that the exponents of power-law part and exponential tapering, and in most cases even the characteristic mass scale of the tapering, remain roughly constant in time after the dust self-gravity has reached its final strength. The evolution of the mass distribution in the simulation excluding sink particle mergers that is depicted in Figure~\ref{fig:IMF_time_series} corroborates this finding. That is, the planetesimal ``birth'' mass distribution can be measured just as well at the end of the simulations as earlier. On the other hand, the figure also shows that the shape of the mass distribution in the simulation including mergers does not remain constant. However, as the shape in this simulation begins to evolve already while the strength of self-gravity is still raised, we cannot determine an appropriate time to measure the birth mass distribution in this model.

When fitting Equation~\ref{eq:tapered_power_law} to the cumulative mass distributions in our simulations\footnote{Even though thousands of planetesimals form in our simulations, we were not able to obtain fits to the differential mass distributions that are as robust as the ones to the cumulative mass distributions, due to the increased noise level in the former. We therefore investigate only fits to the latter.}, we applied the method of least squares and presumed that the uncertainty in the cumulative numbers~$N_>$ is proportional to~$N_>$. This results in more accurate fits at the high-mass end at the expense of the low-mass end, reflecting the higher confidence that the numbers at high masses are complete. In addition, we considered the uncertainty to scale with~$M^{\gamma}$. To measure the goodness of our fits, we conducted chi-squared tests, with~$\chi^2$ thus being given by
\begin{equation}
\chi^2=\sum_M\frac{\left(N_{>,\textrm{sim}}(M)-N_{>,\textrm{fit}}(M)\right)^2}{N_{>,\textrm{sim}}(M)\,M^{\gamma}}.
\label{eq:goodness_of_fit_parameter}
\end{equation}
Here,~$N_{>,\textrm{sim}}$ and~$N_{>,\textrm{fit}}$ are the cumulative numbers arising in the simulations and the fitted numbers, respectively.

\begin{figure*}[t]
  \begin{minipage}{0.49\textwidth}
    \centering
    \includegraphics[width=\textwidth]{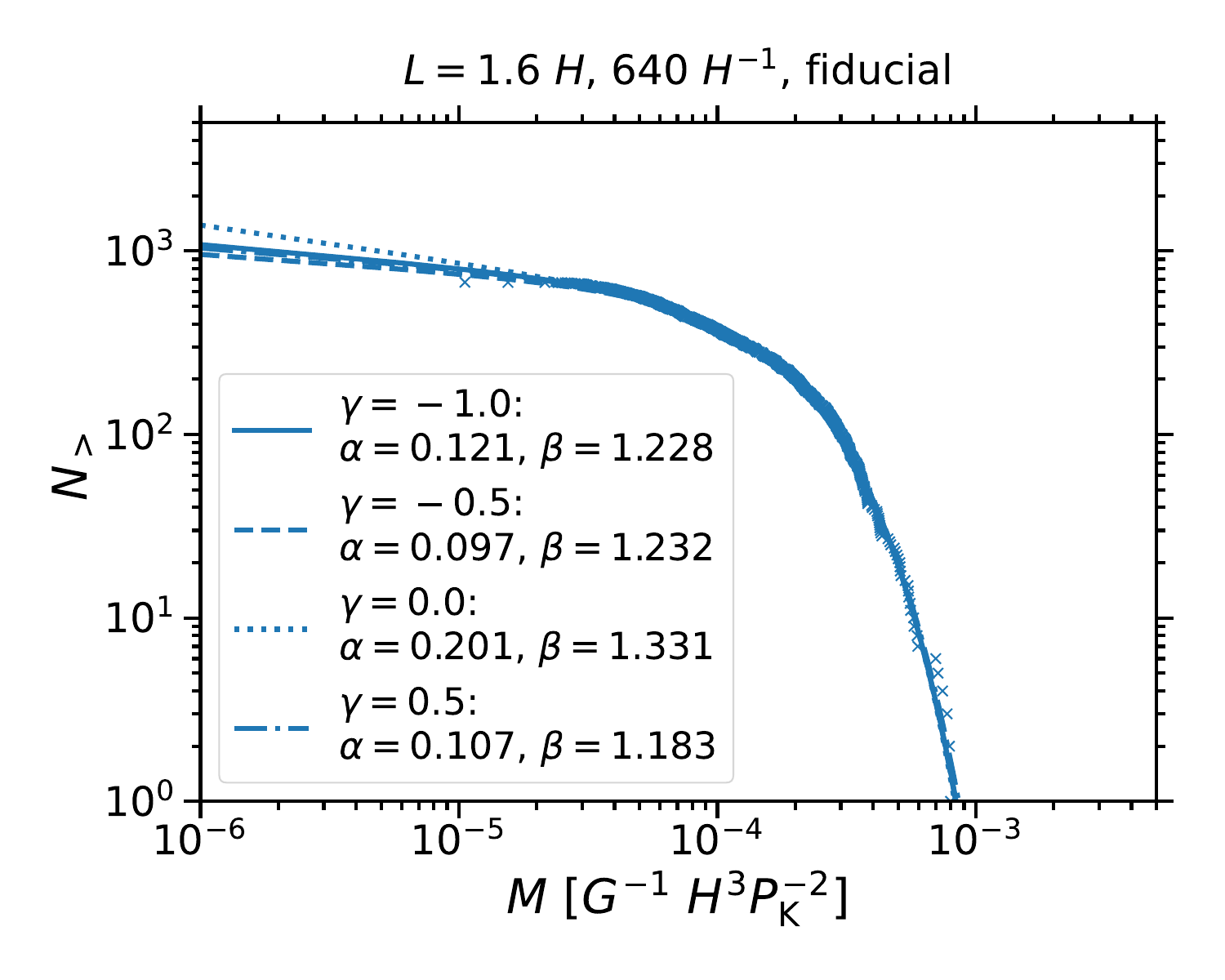} 
  \end{minipage}
  \hfill
  \begin{minipage}{0.49\textwidth}
    \centering
    \includegraphics[width=\textwidth]{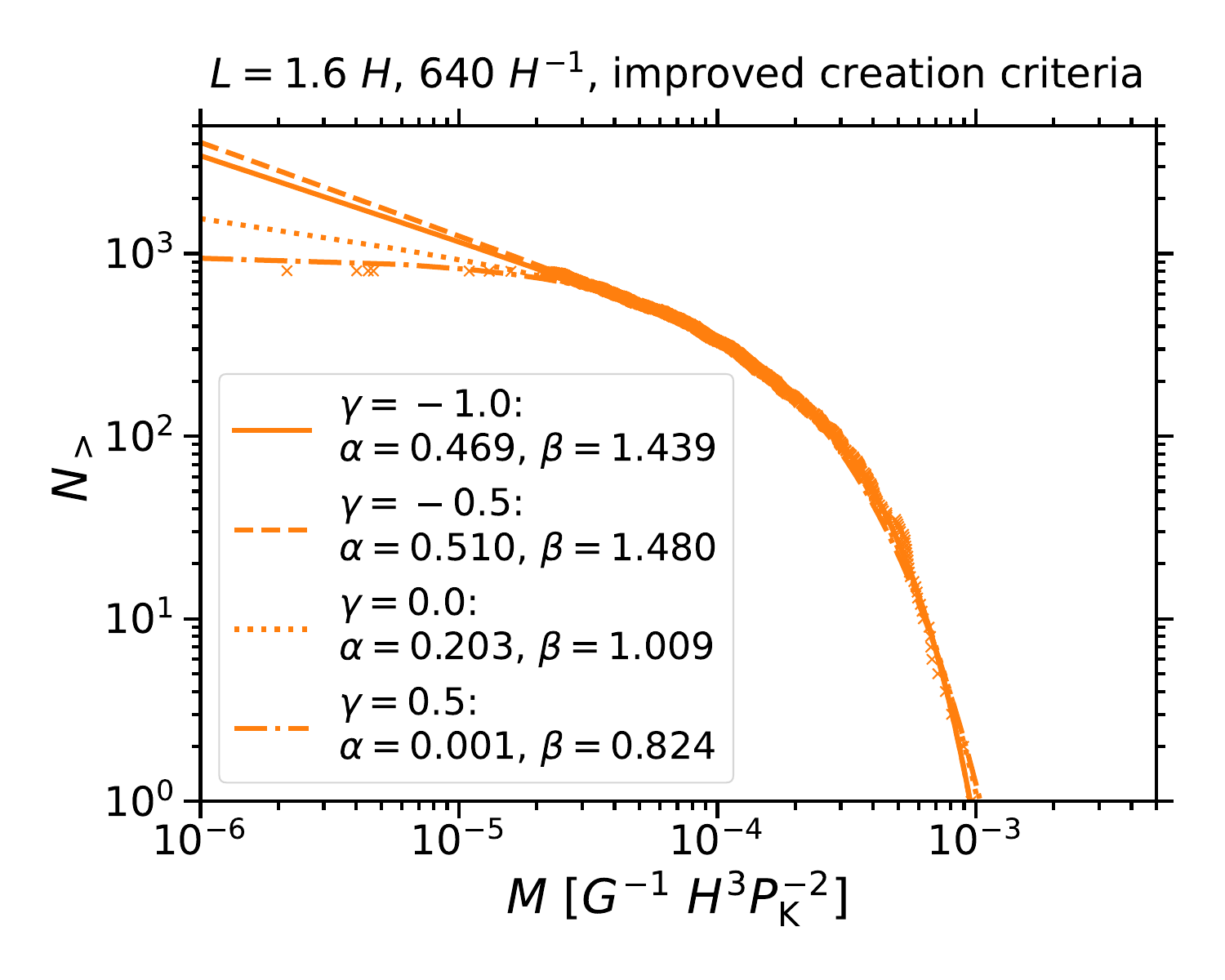} 
  \end{minipage} 
  \caption{Cumulative mass distributions at the end of simulations with the fiducial (left panel) and the improved sink particle creation criteria (right panel) but excluding mutual sink particle accretion. The mass distributions are fitted with an exponentially tapered power law (see Eq.~\ref{eq:tapered_power_law}). The uncertainty of the cumulative numbers~$N_>(M)$ is assumed to be proportional to~$N_>(M)M^{\gamma}$, with fits for different choices of~$\gamma$ being shown. While for the fiducial creation criteria~\mbox{$\gamma=0$} yields the fit with the largest exponents of power-law part~$\alpha$ and exponential tapering~$\beta$, the two exponents are highest if~\mbox{$\gamma=-0.5$} for the improved creation criteria. In both cases, this comparatively steep power law appears to fit the numbers at intermediate masses best.}
  \label{fig:IMF_gamma}
\end{figure*}

Figure~\ref{fig:IMF_gamma} illustrates how different choices of the parameter~$\gamma$ in the above equation affect the fitting to the mass distributions in simulations with the fiducial and the improved sink particle creation criteria, respectively, but without mutual sink particle accretion. In the former case, choosing~\mbox{$\gamma=0$} results in the steepest power law that seems the least constrained by the numbers at low masses and to provide the best fit at intermediate masses. In the latter case, on the other hand, the same is accomplished by setting~\mbox{$\gamma=-0.5$} or~\mbox{$\gamma=-1$} -- that is, assuming a higher uncertainty at low masses relative to at high masses.

Consequently, we selected~\mbox{$\gamma=-0.5$} to fit the mass distributions in our simulations with the improved sink particle creation criteria, and~\mbox{$\gamma=0$} otherwise. In particular, we did not choose a negative value of~$\gamma$ for the distribution obtained by using the clump finding algorithm and the one in the simulation including sink particle mergers since this would put even more emphasis on accurately fitting the high-mass end of these two distributions, which are likely influenced by overly large clumps and artificial mergers, respectively.

\subsubsection{Results}
\label{sect:IMF_results}
\begin{figure}[t]
  \centering
  \includegraphics[width=\columnwidth]{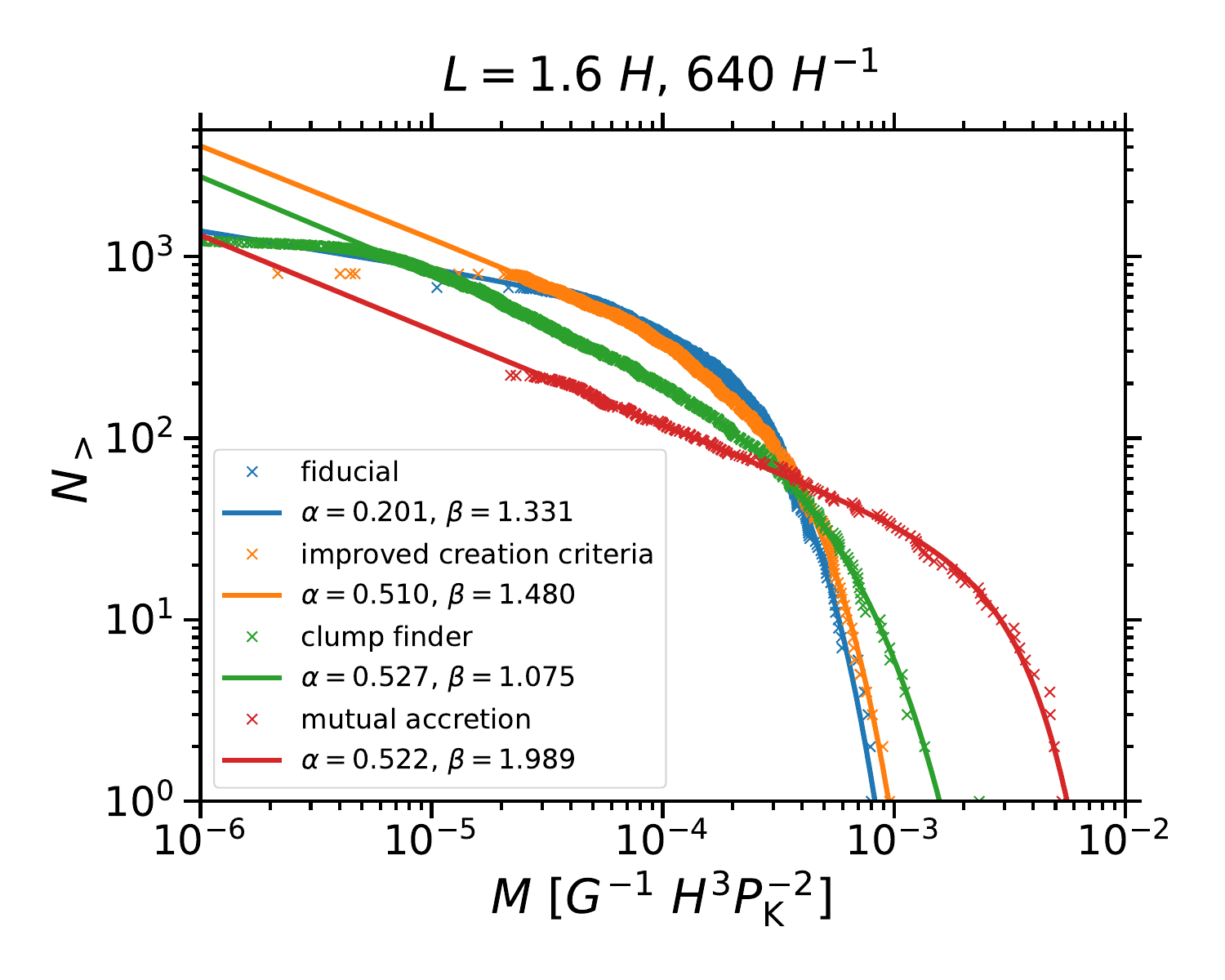} 
  \caption{Cumulative mass distributions of the sink particles or the clumps identified by the clump finding algorithm in simulations with a domain extent and resolution of~\mbox{$L=1.6~H$} and~$640~H^{-1}$, respectively. All distributions are fitted with an exponentially tapered power law (see Eq.~\ref{eq:tapered_power_law}). The power-law parts are comparably steep, with~\mbox{$\alpha=0.5$}, for all simulations apart from the one of the fiducial model. The exponent of the tapering, on the other hand, varies between~\mbox{$\beta=1$} for the clump mass distribution and~$2$ for the sink particle mass distribution in the simulation including sink particle mergers.}
  \label{fig:IMF_fit}
\end{figure}

\begin{table*}[t]
\caption{Best-fitting parameters}
\label{table:IMF_parameters}
\resizebox{\hsize}{!}{
\begin{tabular}{lcccc}
\hline
\hline
Simulation&$\alpha$&$\beta$&$M_{\textrm{pow}}$&$M_{\textrm{exp}}$\\
&&&[$G^{-1}\,H^{3}P_{\rm K}^{-2}$]&[$G^{-1}\,H^{3}P_{\rm K}^{-2}$]\\
\hline
\textit{run\_1.6\_640}&$0.201\pm0.014$&$1.331\pm0.018$&$(3.53\pm0.10)\times10^{-5}$&$(2.19\pm0.05)\times10^{-4}$\\
\textit{run\_1.6\_640}, improved creation criteria&$0.510\pm0.011$&$1.480\pm0.018$&$(2.39\pm0.04)\times10^{-5}$&$(3.28\pm0.05)\times10^{-4}$\\
\textit{run\_1.6\_640}, clump finder\tablefootmark{a}&$0.527\pm0.003$&$1.075\pm0.013$&$(5.98\pm0.03)\times10^{-6}$&$(4.30\pm0.06)\times10^{-4}$\\
\textit{run\_1.6\_640}, mutual accretion&$0.522\pm0.005$&$1.989\pm0.060$&$(2.99\pm0.04)\times10^{-5}$&$(3.43\pm0.05)\times10^{-3}$\\
\textit{run\_3.2\_640}, improved creation criteria&$0.394\pm0.009$&$1.213\pm0.010$&$(2.76\pm0.04)\times10^{-5}$&$(2.61\pm0.04)\times10^{-4}$\\
\end{tabular}
}
\tablefoot{
Listed errors are standard errors.
\tablefoottext{a}{Only the mass distribution of clumps more massive than~$5\times10^{-6}\,G^{-1}\,H^3P_{\rm K}^{-2}$ was fit.}
}
\end{table*}

Figure~\ref{fig:IMF_fit} depicts the exponentially tapered power laws that constitute the best fits to the planetesimal mass distributions in our simulations with a domain size of~\mbox{$L=L_x=L_y=1.6~H$} and a resolution of~$640~H^{-1}$. We show both the sink particle mass distributions and the mass distribution established using the clump finding algorithm. The best-fitting parameters are listed in Table~\ref{table:IMF_parameters}, which additionally includes a simulation with~\mbox{$L=3.2~H$}.

\begin{figure*}[t]
  \centering
  \includegraphics[width=\textwidth]{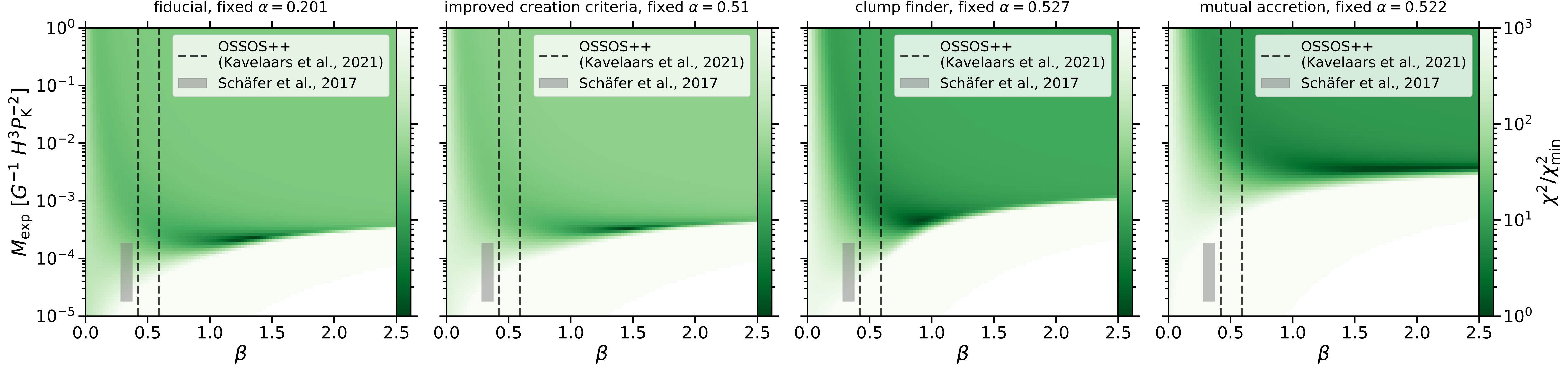}
  \vspace{0em}\\
  \includegraphics[width=\textwidth]{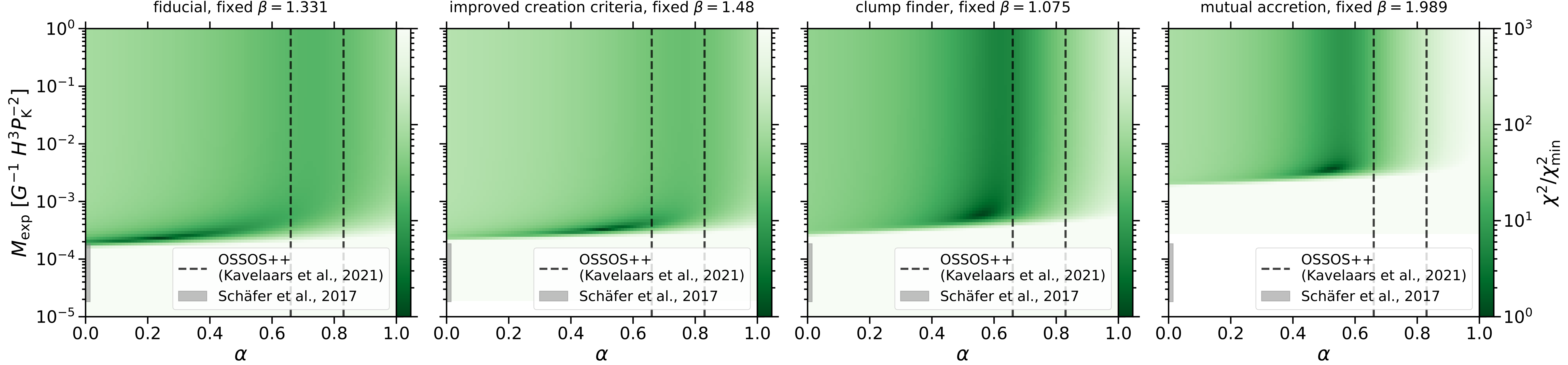}
  \caption{Goodness-of-fit parameter~$\chi^2$ as a function of the exponent~$\beta$ and characteristic mass of the exponential tapering~$M_{\textrm{exp}}$ (top panels) as well as of the exponent of the power-law part~$\alpha$ and the characteristic tapering mass (bottom panels). The different panels depict the parameters of exponentially tapered power laws fit to the cumulative mass distributions of the sink particles in different models or of the clump mass distribution identified by the clump finder, respectively. The goodness-of-fit parameter is defined in Eq.~\ref{eq:goodness_of_fit_parameter} and normalised to the lowest value for the respective model. Tapering exponents between~$1$ and~$1.5$ are preferred for the sink particle mass distributions in models without sink particle mergers (first and second panel), a value close to~$1$ for the clump mass distribution (third panel), and values between~$1$ and~$2.5$ or even higher for our model including mergers (fourth panel). Furthermore, power-law exponents ranging from~$0.4$ to~$0.6$ yield low goodness-of-fit parameter values for all models, with the values for exponents down to~$0$ being comparably low for the model with the fiducial sink particle creation criteria and excluding mergers (first panel). Lastly, characteristic tapering masses of the order of~$10^{-3}~G^{-1}\,H^{3}P_{\textrm{K}}^{-2}$ provide the best fits for the model with mergers, and of the order of~$10^{-4}~G^{-1}\,H^{3}P_{\textrm{K}}^{-2}$ for the other three models.}
  \label{fig:IMF_alpha_beta}
\end{figure*}

In addition, we show in Figure~\ref{fig:IMF_alpha_beta} the goodness-of-fit parameters~$\chi^2$ computed according to Eq.~\ref{eq:goodness_of_fit_parameter} for a broad range of combinations of the exponent and the characteristic mass scale of the tapering as well as of the power-law exponent and the tapering mass scale. In both cases, we fix the respective other exponent and the power-law mass scale to their best-fitting values.

Despite the resolution being limited, the power-law part of the initial mass functions is only marginally shallower than in previously presented models with higher resolutions where~\mbox{$\alpha=0.6$} \citep{Johansen+etal2015, Simon+etal2016, Simon+etal2017, RucskaWadsley2021}. As can be seen from Figure~\ref{fig:IMF_alpha_beta}, power-law exponents ranging from~$0.4$ to~$0.6$ are preferred in three of the four models. In line with this result, \citet{Abod+etal2019} find the initial mass function to be well-described by solely a power law with an exponent of~$0.6$ but fitted even better by an exponentially tapered power law with a smaller value of this exponent. Nonetheless, an exponent of~$0.3$ yields the best fit to the distributions in their models, in comparison with values between~$0.4$ and~$0.5$ for the three models (see Figure~\ref{fig:IMF_fit} and Table~\ref{table:IMF_parameters}). The exception here is our fiducial model, for which values between~$0$ -- that is to say, the mass distribution being fitted with only an exponential tapering -- and~$0.5$ yield low goodness-of-fit parameter values, with~$0.2$ providing the best fit. 

On the other hand, the exponential taperings of the mass distributions in this study are considerably steeper, with best-fitting exponents of~\mbox{$\beta=1-2$}, than those of the distributions in our previous study \citep{Schafer+etal2017}, where we measure exponents ranging between~$0.3$ and~$0.4$. Furthermore, comparatively high values of the goodness-of-fit parameter result from the preferred combinations of exponent and characteristic mass of the tapering established in \citeauthor{Schafer+etal2017} (\citeyear{Schafer+etal2017}; see Figure~\ref{fig:IMF_alpha_beta}). 

Likely, the discrepancies between this work and  \citet{Schafer+etal2017} arise from the domains simulated in the latter work being both smaller (only up to~\mbox{$L=0.8~H$}) and less or equally well resolved (either~$320~H^{-1}$ or~$640~H^{-1}$). On the one hand, we detail above that in domains with azimuthal sizes similar to or smaller than~$1~H$, the streaming-instability-induced dust filaments appear axisymmetric, potentially resulting in planetesimals accreting dust from an overly large feeding zone and thus growing overly massive. On the other hand, we show in the right panel of Figure~\ref{fig:IMF_domain_size_resolution} that a tapering exponent of~$0.6$ -- more consistent with the exponents found by \citet{Schafer+etal2017} -- yields the best fit to the mass distribution in our simulation with~\mbox{$L=0.8~H$} and a resolution of~$320~H^{-1}$, compared with an exponent of~$1.8$ for the distributions in the simulations with the same domain extent and resolutions of~$640~H^{-1}$ or~$1280~H^{-1}$.

As shown in Figure~\ref{fig:IMF_alpha_beta}, a tapering exponent of~$1$ provides a good fit to the mass distributions in all our models, and the best fit to the distribution obtained employing the clump finding algorithm. The sink particle mass distributions in models including or excluding mergers are similarly well-fitted with exponents of up to~$1.5$ or up to~$2.5$ and larger, respectively. These values are in agreement with the exponents of~$1$ and~$4/3$ established by \citet{Abod+etal2019} and \citet{Johansen+etal2015}. We note, though, that these authors assume a fixed exponent rather than treating it as a fitting parameter.

Finally, the characteristic mass of the exponential tapering~$M_{\textrm{exp}}$ is generally of the order of~$10^{-4}~G^{-1}\,H^{3}P_{\textrm{K}}^{-2}$. To compare to the classical Kuiper belt objects as an example, we can convert this mass from code units to natural units by assuming a radius of~$40~\au$ in the passive disc \citep{ChiangGoldreich1997} around a Solar-mass star. Under these assumptions, the characteristic tapering mass corresponds to~$1.763\times10^{24}~\g$ or about half of the mass of the dwarf planet Makemake. Here, our model including mutual sink particle accretion is an exception since probably artificial mergers enhance the characteristic mass by an order of magnitude.

\section{Discussion}
\label{sect:discussion}
\subsection{Numerically measuring the planetesimal initial mass function is challenging}
Our study highlights a number of challenges that arise when employing numerical models of the streaming instability to obtain the planetesimal initial mass function.

Firstly, because computational resources are limited, trade-offs between simulation domain size and resolution are unavoidable. We argue in Sects.~\ref{sect:filaments} and~\ref{sect:domain_size_resolution} that to reach convergence with respect to the shape of the initial mass function, domains with an azimuthal extent of one gas scale height or more are necessary. This is because in smaller domains with periodic boundaries the dust filaments caused by the streaming instability appear axisymmetric although they in fact are not. Consequently, the feeding zone of planetesimals emerging from these filaments is boundless in the azimuthal dimension, resulting in artificially enhanced planetesimal masses. Nevertheless, in previous work on the initial mass function, with the exception of our previous study \citep{Schafer+etal2017}, exclusively domains with an azimuthal size of no more than~$0.2$ scale heights have been employed.

On the other hand, less massive planetesimals form in simulations with higher resolutions \citep{Johansen+etal2015, Simon+etal2016, Li+etal2019}. Highly resolved models are therefore required to put strong constraints on the power-law part of the initial mass function; and on the turnover at low masses that was discovered by \citet{Li+etal2019} in their simulation with a resolution of~$5120$ cells per gas scale height, the highest to date. In contrast, despite utilising 40 million CPU hours to conduct the simulations presented in this study, we could not afford to resolve our domains with azimuthal sizes greater than one scale height with more than~$640$ cells per scale height. (A simulation of ours with an extent of~$1.6~H$ in the plane and a resolution of~$640~H^{-1}$ consumed~\mbox{${\sim}0.3$} million CPU hours. Doubling the resolution increases the computational cost by roughly a factor of~$16$, a factor of~$8$ owing to the greater number of grid cells and a factor of~$2$ because of the shorter time step.)

Furthermore, whereas authors applying Athena \citep[e.g.][]{Simon+etal2016, Simon+etal2017, Li+etal2019, RucskaWadsley2021} use clump finding algorithms to identify planetesimals and measure their masses, studies like ours employing the Pencil Code \citep{Johansen+etal2015, Schafer+etal2017} involve sink particles for this purpose. While the former method presents its own challenges, the latter one necessitates carefully considering approaches to sink particle creation and accretion. We added new creation criteria that are detailed in Sect.~\ref{sect:sink_particles} to the Pencil Code, but do not find the choice between the previously implemented criteria and these improved ones to significantly affect the planetesimal mass distribution.

Nonetheless, multiple sink particles can be created in the same gravitationally unstable dust clump both under the old and the new criteria. Since we do not resolve binary planetesimals or multiples, it would be desirable for these sink particles to merge until eventually only one remains. We therefore performed a simulation including mutual sink particle accretion. However, as we discuss in Sect.~\ref{sect:accretion}, we find very few mergers in this simulation to likely be of sink particles originating from the same clump. More importantly, sink particle mergers are not resolved, with the maximum impact parameter leading to a collision of two sink particles typically amounting to only a few percent of the cell size.

More sophisticated sink particle algorithms could be taken into consideration in future studies. For instance, we chose a dust density threshold for sink particle creation based on the Roche density -- the choice of this threshold is problematic in itself because too low a threshold results in the creation of sink particles before self-gravity is even introduced in our model (see Sect.~\ref{sect:accretion}), while too high a threshold suppresses creation even if it would be physical. Nevertheless, in addition to the stellar tidal forces that determine the Roche density, the tidal forces exerted by already existing sink particles could be taken into account. Further inspiration can be gained from the creation criteria applied in models of star formation \citep[e.g.][]{Federrath+etal2010, GongOstriker2013, Haugbolle+etal2018}. Moreover, an approach to mutual sink particle accretion that goes beyond merging sink particles located in the same cell would be of interest, potentially employing sub-grid models.

Lastly, as evident from Figure~\ref{fig:IMF_alpha_beta}, even though the thousands of planetesimals forming in our large simulation domains provide robust statistics for fitting the planetesimal mass distribution, the fitting parameters are subject to considerable degeneracy. It is therefore beneficial to not only report best-fitting parameters, but ranges of parameter values that yield good fits to the mass distribution. 

\subsection{Comparison with the cold classical Kuiper belt object mass distribution}
In previous studies of the streaming instability, the planetesimal mass distribution has mostly been described using either a simple power law \citep[e.g.][]{Simon+etal2016, Simon+etal2017, Abod+etal2019, RucskaWadsley2021} or a power law with an exponential tapering \citep{Johansen+etal2015, Schafer+etal2017, Abod+etal2019}, with \citet{Li+etal2019} showing that multi-segment power laws provide the best fits to the mass distributions in their simulations.

In this work, we focus on the exponentially tapered power law expressly because this permits us to compare to the absolute magnitude distribution of the cold classical Kuiper belt objects, which is found to possess this functional form by \citet{Kavelaars+etal2021} and \citet{Napier+etal2024}. Such a comparison is pertinent since these objects are believed to have formed in situ and to have undergone minimal collisional evolution, and thus constitute a largely pristine sample of planetesimals formed in the young Solar System.

\citet{Kavelaars+etal2021} present two fits to the absolute magnitude distribution of the debiased OSSOS++ sample of cold classical Kuiper belt objects. These provide reasonably good fits also to the planetesimal mass distributions arising in our models. Fixing the power-law exponent at either~\mbox{$\alpha=0.66$} or~$0.83$, these authors obtain a tapering exponent of~\mbox{$\beta=0.42$} or~$0.59$, respectively. That is, by tendency the power-law part is steeper but the tapering shallower than those of the planetesimal mass distributions in our models (see Table~\ref{table:IMF_parameters} and Figure~\ref{fig:IMF_fit}). Still, as can be gathered from Figure~\ref{fig:IMF_alpha_beta}, these combinations of exponents can yield low goodness-of-fit parameter values (depending on the choice of model and characteristic mass scale of the tapering). We note that \citet{Petit+etal2023} find the absolute magnitude distributions of hot and cold classical Kuiper belt objects to overall be very similar, but only the latter to possess an exponential tapering.

Converting the masses of the planetesimals in our simulations to absolute magnitudes in order to compare to the cold classical Kuiper belt objects requires a number of assumptions. Placing the simulation domains at~$40~\au$ -- roughly the location of the cold classical Kuiper belt -- in a passive disc \citep{ChiangGoldreich1997} around a Solar-mass star and presuming spherical planetesimals with a density of~$0.5~\g\,\cm^{-3}$ appropriate for icy bodies as well as an albedo of~$0.5$ results in absolute magnitudes that are three magnitudes larger than those of the cold classical Kuiper belt objects.

However, we note that the initial conditions of our simulations were chosen to ensure that the streaming instability causes filament and planetesimal formation rather than to reflect the conditions in the early Solar System. Specifically, based on previous work \citet{Liu+etal2020} establish that the characteristic planetesimal mass scales with~$\gamma^{1.5}$, where~$\gamma$ is the dimensionless parameter quantifying the strength of the dust self-gravity. That is, reducing the rather arbitrarily chosen value of this parameter by two orders of magnitude would result in planetesimal masses that are three orders of magnitude smaller and thus in accordance with the masses of the cold classical Kuiper belt objects. This illustrates the benefit of future studies with significantly lower self-gravity parameters. Nonetheless, it is important to note that the shape of the planetesimal mass distribution is found to be independent of the choice of self-gravity parameter by \citet{Simon+etal2016}.

\section{Conclusion}
\label{sect:conclusion}
We present numerical simulations of the accumulation of dust in dense filaments owing to the streaming instability, and of the formation of planetesimals from gravitationally unstable clumps in these filaments. The domains of our simulations span up to~$6.4$ gas scale heights in the radial and azimuthal dimensions and are thus larger by a factor of~$32$ in both dimensions and by almost three orders of magnitude in volume than the domains that are typically considered for these kinds of simulations. Consequently, an unprecedentedly large sample of planetesimals emerges in our simulations, providing us with robust statistics to constrain the shape of the high-mass end of the planetesimal initial mass function.

A key result of our study is that the dust filaments induced by the streaming instability possess azimuthal sizes of no more than about one gas scale height. Simulating domains with periodic boundary conditions that are smaller than this size therefore leads to exceedingly high planetesimal masses. This is since filaments seem axisymmetric in such domains, and the feeding zone of planetesimals that form and accrete from the filaments thus appear to be unbounded in the azimuthal dimension.

The initial mass function of the planetesimals forming in our models is well-represented by a power law with a steep exponential tapering at the highest masses. We can only afford to apply a resolution of~$640$ grid cells per gas scale height to our large simulation domains -- low compared to the resolutions employed in previous studies of smaller domains -- and the planetesimal numbers are therefore incomplete at low masses \citep{Johansen+etal2015, Simon+etal2016, Li+etal2019}. Nonetheless, when weighing the numbers at intermediate and high masses more heavily than these, both the exponents of the power-law part and of the tapering we find are consistent with the values obtained in previous work \citep{Johansen+etal2015, Simon+etal2016, Simon+etal2017, Abod+etal2019, RucskaWadsley2021}. Both the power-law and the tapering exponent can be reconciled with the values which \citet{Kavelaars+etal2021} measure for the absolute magnitude distribution of the cold classical Kuiper belt objects when considering degeneracies in the parameters of the exponentially tapered power law and numerical uncertainties, including those arising from the approach to modelling the observational properties of planetesimals such as surface albedo.

\begin{acknowledgements}
We are grateful to the anonymous referee for their comments
that helped to improve this paper. To analyse and visualise the simulations presented in this paper, the Python libraries Matplotlib\footnote{\url{https://matplotlib.org}} \citep{Hunter2007}, NumPy\footnote{\url{https://numpy.org}} \citep{Harris+etal2020}, and SciPy\footnote{\url{https://scipy.org}}\citep{Virtanen+etal2020} were used. We acknowledge PRACE for awarding us a grant of 40 million CPU hours at Joliot-Curie, GENCI@CEA, France. U.S. and A.J. are thankful for funding from the Danish National Research Foundation (DNRF Chair Grant DNRF159). A.J. further gratefully acknowledges funding from the Knut and Alice Wallenberg Foundation (Wallenberg Scholar Grant 2019.0442), the G\"oran Gustafsson Foundation, and the Carlsberg Foundation (Semper Ardens: Advance grant FIRSTATMO). T.H. acknowledges funding from the Independent Research Fund Denmark through grant No. DFF 8021- 00350B.
\end{acknowledgements}

\bibliography{references}

\begin{appendix}

\section{Dependence of planetesimal initial mass function on dust Stokes number and strength of gas pressure gradient}
\label{sect:appendix}

\begin{figure}[t]
  \centering
  \includegraphics[width=\columnwidth]{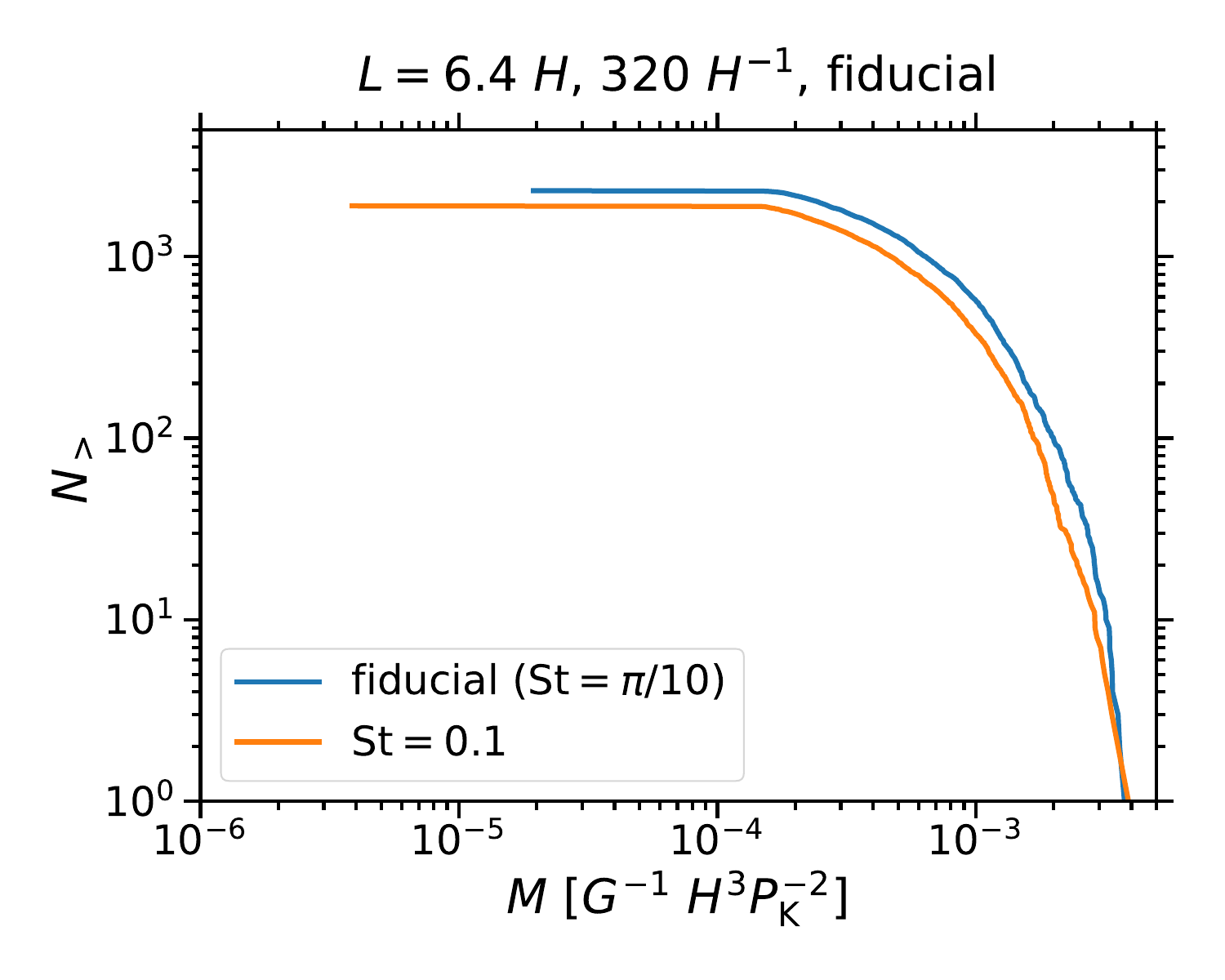} 
\caption{Cumulative mass distributions of the sink particles at the end of two simulations of our fiducial model with dust Stokes numbers of~\mbox{$\St=0.314$} (blue line) and~$0.1$ (orange line). The shapes of the mass distributions are comparable, though the sink particle number and masses are higher in the former simulation.}
\label{fig:IMF_Stokes_number}
\end{figure}

Figure~\ref{fig:IMF_Stokes_number} depicts the planetesimal mass distributions in two simulations with different Stokes numbers of the dust. It is evident that the shape of the distributions is similar, although overall somewhat more and more massive planetesimals emerge in the simulation with the larger Stokes number. In agreement with these results, \citet{Simon+etal2017} show that the planetesimal initial mass function is well-fitted with a power law with an exponent of~$\alpha=0.6$ independent of the Stokes number, even though the planetesimal number and masses increase with the Stokes number. The latter is consistent with the streaming instability giving rise to stronger dust concentration for higher Stokes numbers \citep[see Figure~\ref{fig:filaments_Stokes_number};][]{YoudinGoodman2005, JohansenYoudin2007, Carrera+etal2015}.

\begin{figure}[t]
  \centering
  \includegraphics[width=\columnwidth]{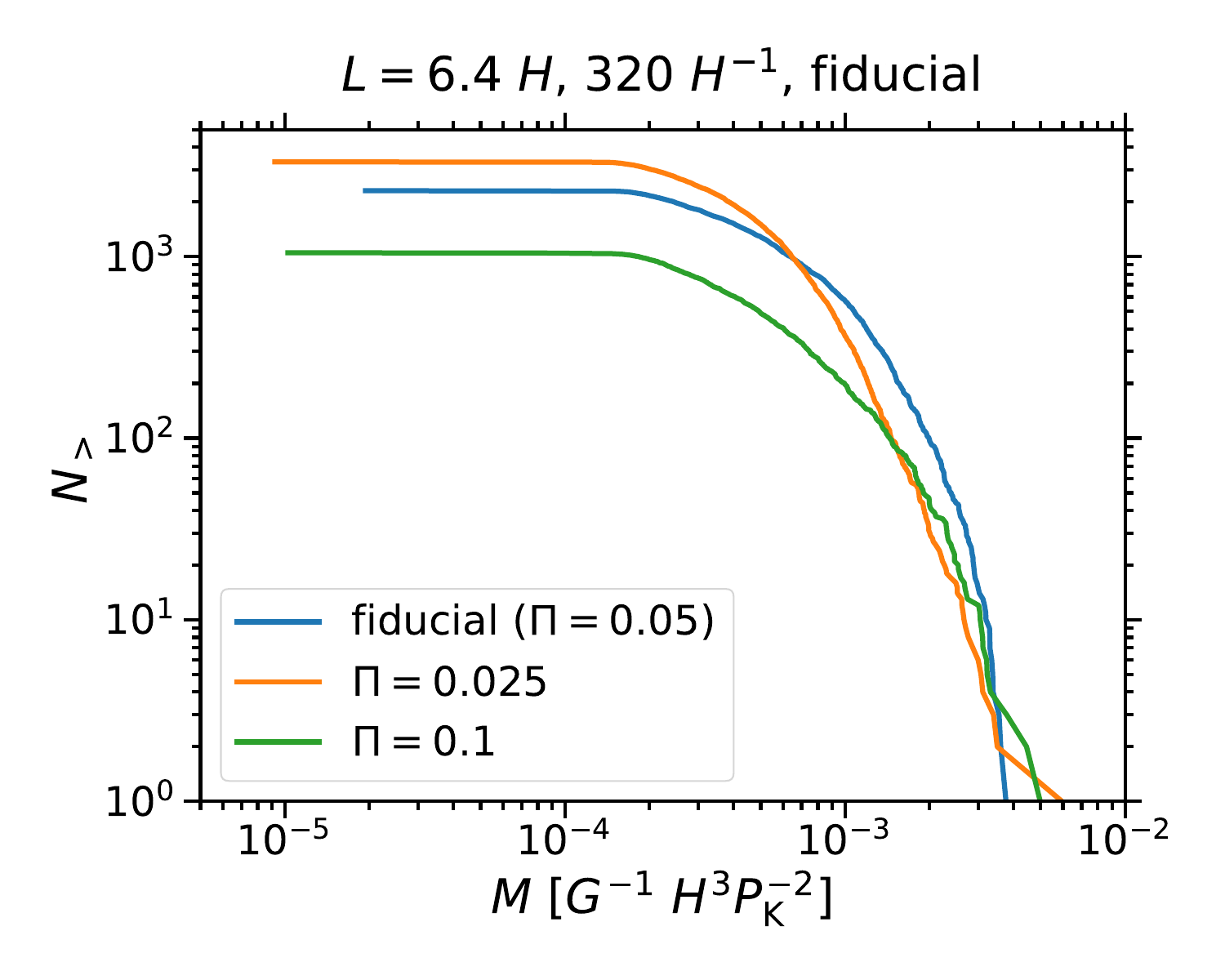} 
\caption{Cumulative sink particle mass distributions at the end of three simulations of our fiducial model with varying strengths of the gas pressure gradient. While the sink particles are less and less numerous with increasing gradient strength, their minimum and maximum mass are similar in all three simulations. The overall steepness of the distribution thus decreases for stronger gradients.}
\label{fig:IMF_pres_grad}
\end{figure}

The planetesimal mass distribution possesses a more complex dependence on the strength of the gas pressure gradient, as can be seen from Figure~\ref{fig:IMF_pres_grad}. While the minimum and maximum planetesimal masses remain largely constant, fewer planetesimals form in models in which the pressure gradient is stronger. This leads to the distribution on the whole being shallower. In line with these findings, \citet{Abod+etal2019} measure a lower power-law exponent of the planetesimal initial mass function for higher pressure gradient strengths. In their model, both the number and the maximum mass by tendency decline with the strength, though (see their Figure~6).

\end{appendix}
\end{document}